\documentstyle[11pt, amsfonts]{article}

\begin{document}

\def\vac{| 0 >}
\def\leftvac{< 0 |}
\def\C{\mathbb C}
\def\R{\mathbb R}
\def\S{\mathbb S}

\begin{center}

{\Large {\bf ``QUANTIZATION IS A MYSTERY''}}\footnote{First part of a famous
aphorism of Edward Nelson that ends with ``{\it but second quantization is a
functor}''. To quote John Baez \cite{B06} ``No one is a true mathematical
physicist unless he can explain'' this saying.

{\it Bulg. J. Phys.} {\bf 39} (2012) 107-149;
Bures-sur-Yvette preprint IHES/P/12/01.}

\vspace{4mm}

Ivan Todorov

Institut des Hautes \'Etudes Scientifiques, F-91440 Bures-sur-Yvette, France

and

Theory Division, Department of Physics, CERN, CH-1211 Geneva 23, Switzerland

Permanent address:
 
Institute for Nuclear Research and Nuclear Energy\\
Tsarigradsko Chaussee 72, BG-1784 Sofia, Bulgaria\\
e-mail: todorov@inrne.bas.bg 

\end{center}

\begin{abstract}

Expository notes which combine a historical survey of the development of quantum physics with a review of selected mathematical topics in quantization theory (addressed to students that are not complete novices in quantum mechanics).

After recalling in the introduction the early stages of the quantum revolution, and recapitulating 
in Sect. 2.1 some basic notions of symplectic geometry, we survey in Sect. 2.2 the 
so called {\it prequantization} thus preparing the ground for an outline of 
{\it geometric quantization} (Sect. 2.3). In Sect. 3 we apply the general theory
to the study of basic examples of {\it quantization of K\"ahler manifolds}. In Sect. 4 we review the Weyl and Wigner maps and the work of Groenewold and Moyal that laid the foundations of {\it quantum mechanics in phase space}, ending with a brief survey of the modern development of {\it deformation quantization}. Sect. 5
provides a review of {\it second quantization} and its mathematical interpretation. We point out that the treatment of (nonrelativistic) bound states requires going beyond the neat mathematical formalization of the concept of second quantization. An appendix is devoted to Pascual Jordan, the least known among the creators of  quantum mechanics and  the chief architect of the ``theory of quantized matter waves''.
\end{abstract}

\bigskip


\newpage

\tableofcontents
\bigskip

\noindent {\bf Appendix A \ \ Pascual Jordan (1902-1980) \hfill 38}
 
\newpage

\section{Introduction: historical remarks}
\setcounter{equation}{0}
\renewcommand\theequation{\thesection.\arabic{equation}}

\smallskip
Quantum mechanics - old and new - has been an active subject for nearly a century. Even if we only count  textbooks the number is enormous - and keeps growing. My favourite is Dirac's \cite{D30}. These notes are addressed to readers with a taste in the history of the subject and in its mathematical foundations. An early monograph on the mathematical meaning of quantum mechanics is John von Neumann's \cite{vN}. For more recent texts - see \cite{FY, Mac, T} among many others. The latter book also contains a well selected bibliography. Sources on the history of the subject include \cite{MR, Dar, Sch, PJ07}.

\subsection{First steps in the quantum revolution}

Quantum theory requires a new conceptual basis. Such a drastic change of the
highly successful classical mechanics and electrodynamics was justified by the
gradual realization at the turn of 19th century that they are inadequate in the
realm of atomic phenomena. Four theoretical breakthroughs prepared the creation
of quantum mechanics.

{\bf 1900}: Following closely the Rubens-Kurlbaum experiments in Berlin Max
Planck (1858-1947) found the formula for the spectral density $\rho(\nu, T)$ of
 the black-body radiation  as a function of the frequency $\nu$ and the
absolute temperature $T$:
\begin{equation}
\label{Planck}
\rho(\nu, T) = \frac{8\pi h\nu^3}{c^3} \frac{1}{e^{\beta h \nu} - 1}\,, \quad \beta = \frac{1}{kT}\,.
\end{equation}
Here $k$ is Boltzmann's\footnote{Ludwig Boltzmann (1844-1906) founded the
statistical interpretation of thermodynamics which Planck originally tried to
overcome.The expression for the entropy in terms of probability $S = k log W$
is carved on Boltzmann's gravestone in Vienna (see \cite{Bl}).} constant, $h$ is the {\it Planck's
constant} representing the {\it quantum of action} (that becomes a hallmark of
all four breakthroughs reviewed here). It looks like a miracle that such a
formula should have been found empirically. At the time of its discovery nobody
 seems to have realized that it is closely related to the well known generating
 function of the Bernoulli\footnote{Jacob Bernoulli (1654-1705) is the first in
 the great family of Basel's mathematicians. The Bernoulli numbers appear in
his treatise {\it Ars Conjectandi} on the theory of probability, published
posthumously, in 1713.} numbers, and, more recently, to modular forms. (For a
derivation based on the theory of free massless quantum fields on conformally
compactified space that emphasizes the relation to modular forms - see
\cite{NT}; it has been also related to index and signature theorems - see e.g.
\cite{H71} Sect. 2.) Planck did not stop at that. He found the prerequisites for
 its validity (wild as they sounded at that time). First he assumed that the
energy consists of finite elements, {\it quanta} proportional to the frequency
$\nu$ of the light wave, $\epsilon = h\nu$. Secondly, he recognized that the
quanta should be {\it indistinguishable}, thus anticipating the {\it
Bose-Einstein statistics}, discovered more than two decades later (see \cite{P}
 Sect. {\bf 19a}).That is how Planck, conservative by nature, started, at the
age of 42, the scientific revolution of the 20th century.

{\bf 1905}: Albert Einstein (1879-1955) was the first to appreciate the
revolutionary character of Planck's work. The light-quantum is real: it may
kick electrons out of a metal surface, thus giving rise to the {\it photoelectric
effect}. One can judge how far ahead of his time Einstein went with his bold
hypothesis by the following comment of Planck et al. who recommended him, in
1913, for membership in the Prussian Academy: ``In sum, ... there is hardly
one among the great problems in which modern physics is so rich, to which
Einstein has not made a remarkable contribution. That he may sometimes have
missed the target, as, for example, in his hypothesis of light-quanta, cannot
really be held too much against him ...'' (\cite{P}, {\bf 19f}). Robert Milikan
 (1868-1953) who confirmed Einstein's prediction in 1915 (Nobel Prize in 1923)
could not bring himself to believe too in the ``particles of light''. Even after
 Einstein was awarded the Nobel Prize in 1921 ``especially for his work on the
photoelectric effect'' leading physicists (like Bohr, Kramers\footnote{The Dutch
physicist Hendrik Anthony (``Hans'') Kramers (1894-1952) was for nearly 10
years, 1916-26, the senior collaborator of Niels Bohr (1885-1962) in Copenhagen.} 
and Slater) continued to feel uncomfortable with the wave-particle duality.

{\bf 1911-13}: As Ernest Rutherford (1871-1937) established in 1911 the
planetary atomic model: light electrons orbiting around a compact, massive,
positively charged nucleus, a highly unstable structure according to the laws
of classical electrodynamics, it became clear that atomic physics requires new
laws. Niels Bohr realized in 1913 that the emission and absorption spectra,
the fingerprints of the atoms (\cite{B05}), can be explained as transitions
between stationary states\footnote{In 1910 the Austrian physicist Arthur Haas
(Brno, 1884 - Chicago,1941) anticipated Bohr's model in his PhD thesis. His
result was originally ridiculed in Vienna. Bohr received the Nobel Prize for 
his model of the atom in 1922.} and he derived Balmer's formula for
 the spectrum of the hydrogen atom (see \cite{P86} {\bf 9}(e)). In the 
 words of the eloquent early textbook on quantum mechanics, \cite{D30}, 
 ``We have here a very striking and general example of the breakdown
 of classical mechanics - not merely an inaccuracy of its laws of motion but
{\it an inadequacy of its concepts to supply us with a description of atomic
events}.'' 

{\bf 1923-24}: Inspired by the coexistence of wave-particle properties of light
quanta, Louis-Victor, prince de Broglie (1892-1987) predicted the wave
properties of all particles. His prediction was confirmed in 1927 by two
independent experiments on electron diffraction. De Broglie was awarded the
Nobel Prize in Physics in 1929.

\smallskip

\subsection{The glorious years: 1925-1932}

\begin{flushright}
{\it Whenever we look back at the development of physical theory
                      in the period between 1925 and 1930
          we feel the joy and the shock of the miraculous.}
                                               Rudolf Haag

\end{flushright}

Quantum mechanics appeared in two guises: Werner Heisenberg (1901-1976) and
Paul Dirac (1902-1984) thought it as a particle theory, Louis de Broglie and
Erwin Schr\"odinger (1887-1961) viewed it as a {\it wave mechanics} \cite{Sch}.
 Although their equivalence was recognized already by Schr\"odinger, only the
{\it transformation theory} provided a general setting for seeing the competing
approaches as different representations/pictures of the same theory. It was
developed chiefly by Pascual Jordan and Dirac (see Appendix).

 In July 1925 a hesitating Heisenberg handed to his G\"ottingen professor Max
Born (1882-1970) the manuscript of a ground breaking paper\footnote{conceived
(after a 7 months stay at Copenhagen) while he was recovering from hay fever on
Helgoland, a tiny island in the North Sea - see \cite{T05}.}  ``Quantum
theoretical reinterpretation of kinematic and mechanical relations'' (for an
English translation with commentary - see \cite{SQM}) - and left for
Leyden and Cambridge. Heisenberg ends his paper with an invitation for ``a
deeper mathematical study of the methods used here rather superficially''. Born
 soon recognized that Heisenberg was dealing without realizing it with matrix
multiplication. He shared his excitement with his former assistant, Wolfgang
Pauli (1900-1958), asking him to work out together the proper mathematical
reformulation of Heisenberg's idea, but Pauli answered in his customary
irreverent style: ``Yes, I know, you are fond
of tedious and complicated formalism. You are only going to spoil Heisenberg's
physical ideas with your futile mathematics.'' (\cite{Sch} p. 8). Only then Born
 made the right choice turning to the 22-year-old Jordan. It was Jordan who,
following the idea of his mentor, first proved what is now called ``the
Heisenberg commutation
relation'', $2\pi i (pq - qp) = h$ - and is carved on the gravestone of Born.
 Dirac, who discovered it independently during the same 1925, related it to the
 Poisson\footnote{Sim\'eon-Denis Poisson (1781-1840) introduced in his {\it 
Trait\'e de m\'ecanique}, 1811, the notion of momentum $p = \frac{\partial T}
{\partial {\dot q}}, T$ being the kinetic energy.} bracket $\{q,p\} = 1$. 
Unlike his friend Pauli, Heisenberg welcomed the development of the apparatus of 
matrix mechanics. Decades later he speaks of the lesson drawn from revealing 
the nature of noncommutative multiplication: ``If one finds a difficulty in a 
calculation which is otherwise quite convincing,
one should not push the difficulty away; one should rather try to make it the
centre of the whole thing.``  (\cite{MR} {\bf 3}, III.1). Before Born-Jordan's
paper was completed he began participating in the work - first with letters to
Jordan from Copenhagen. The collaboration ({\it Dreim\"annerarbeit} - the work
of the three men \cite{BHJ}) was fruitful albeit not easy. Heisenberg believed
that they should start with physically interesting applications rather than
first expanding the apparatus, including the theory of the electromagnetic
field, as Born and Jordan were proposing. He insisted that they just postulate
the {\it canonical commutation relations} (CCR) for a system of $n$ degrees of
freedom (\cite{MR} {\bf 3}, III.1)\footnote{The {\it reduced Planck's constant}
$\hbar$ was introduced by Dirac in his book \cite{D30}.}:
\begin{equation}
\label{CCR}
i[p_j, q_k] = \hbar \delta_{jk} \ \ (\hbar = \frac{h}{2\pi})\,,\ \
[p_j, p_k] = 0 = [q_j, q_k]\,, \ \ j, k = 1, 2, ..., n\,,
\end{equation}
rather than to try to derive them from the Hamiltonian equations of motion,
following the wish of his coauthors. After a quarter of a century, Wigner\footnote{Jeno 
(later Eugene) Wigner (Budapest, 1902 - Princeton, 1995) was awarded the Nobel Prize in 
Physics in 1963.} returned to the question ''Do the equations of motion determine the
quantum mechanical commutation relations?" (Phys. Rev. 1950). It triggered the discovery of 
{\it parastatistics} (by Green, Messiah, Greenberg) and their super Lie-algebraic generalization 
(by Palev). The last section, devoted to radiation theory, was written by
Jordan alone (\cite{MR} {\bf 3}, IV.2). It contains the first quantum mechanical
derivation of Planck's black-body-radiation formula, a topic belonging to the realm of
quantum electrodynamics. Decades later, in 1962, talking to Van der Waerden (the editor
of \cite{SQM}, 1903-1995), Jordan says that this is his single most important
contribution to quantum mechanics, a contribution that remained unknown and
unappreciated.

A week before the appearance of \cite{BHJ}, on January 27, 1926, in the wake of an inspiring vacation, 
Schr\"odinger submitted the first of a series of four papers entitled ``Quantization as an
eigenvalue problem''. Just because his formulation of wave mechanics based on
an ``wave equation'' is looking quite different from the picture, drown by
Heisenberg, Born, Jordan and Dirac, it widened the scope of quantum theory and
made it ultimately more flexible.

\bigskip

\subsection{Beginning of a mathematical understanding}
{\it Mathematicians are like Frenchmen: whatever you tell them they translate into their own language and forthwith it becomes something entirely different.}
\begin{flushright}
 J.W. Goethe (1749-1832)
\end{flushright}

After Dirac discovered the simple relation between commutators and Poisson brackets
(PB) of coordinates and momenta,
\begin{equation}
\label{qp}
[q, p] = i\hbar \{q, p\}\ (= i\hbar),
\end{equation}
it appeared tempting to postulate a similar relation for more general observables (that is, real functions on 
phase space\footnote{The story of the appearance of the concept of phase space in mechanics, 
or rather, {\it the tangled tale of phase space} is told in \cite{N}.}). This leads immediately to an ordering
problem. The simple commutation relation (CR)
\begin{equation}
\label{quadr}
 \frac{1}{2} [q^2, p^2] = i\hbar (qp + pq)
\end{equation}
suggests using suitably symmetrized products\footnote{A systematic completely 
symmetrized ordering (see Sect. 4.1) was introduced in \cite{We} by Hermann Weyl (1885-1955), 
a student of David Hilbert (1862-1943), whose fame as one of the last universal 
mathematicians approaches that of his teacher.}. This indeed allows to fit the simple-minded 
quantization rule in the case of second degree polynomials of $p$ and $q$. For general cubic 
polynomials, $f(p, q), g(p,q)$ (and canonical PB - see Sect. 2.1) one cannot always have a 
relation of the type
\begin{equation}
\label{CRPB}
[f, g] = i \hbar \{f, g\}
\end{equation}
no matter how $f, g$ and the right hand side are ordered.\footnote{The equation
(\ref{CRPB}) does have a solution in terms of vector fields that will be
displayed in Sect. 2.2 below; we shall also explain why this solution is
physically unsatisfactory.}  The property of being quadratic (or linear), on the
 other hand, is not invariant under canonical transformations. ``One cannot
expect to be able to quantize a symplectic manifold without some additional
structure''\cite{GW}. (A general result of this type was established over two
decades after the discovery of quantum mechanics - see \cite{G46, V51}.)
 One can at best select a subset of observables for which (\ref{CRPB}) is valid.
 If the problem admits a continuous symmetry then it is wise to choose its Lie
algebra generators among the selected dynamical variables. The above mentioned
example of (symmetric) quadratic polynomials in $p$ and $q$ is of this type:
for a system of $n$ degrees of freedom these polynomials span the Lie algebra
$sp(2n, \mathbb{R})$ corresponding to a projective representation of the real
symplectic group $Sp(2n, \mathbb{R})$ that is a true representation of its
double cover\footnote{That is not a matrix group; more about $Mp(2n)$ and its applications 
can be found in the monographs \cite{F, deG} as well as in Secs. 3 and 4 of 
\cite{T10} and references therein.}, the {\it metaplectic
group} $Mp(2n)$, the authomorphism group of the CCR (\ref{CCR}). It is a
noncompact simple Lie group whose nontrivial unitary irreducible representations
(UIR) are all infinite dimensional. Another physically important example,
considered by Jordan and Heisenberg in \cite{BHJ} is the angular momentum -
the hermitean generators of the Lie algebra $so(3)$ of the (compact) rotation
group (and of its two-fold cover $SU(2)$ that gives room to a half-integer spin\footnote{The story of spin 
is told in \cite{Tom}; for its relation to Clifford algebras - see \cite{T11}.} ${\bf s}$):
\begin{equation}
\label{M}
{\bf M} = {\bf r}\times {\bf p} + {\bf s} \,, \  \ [M_x, M_y] = i\hbar M_z \ \ {\rm etc.} \ \ (M_z = xp_y - yp_x +s_z).
\end{equation}
The following elementary exercise recalls how the representation theory of compact Lie groups
and the CR (\ref{M}) can be used to compute the joint spectrum of $M_z$ and ${\bf M}^2 :=
M_x^2 + M_y^2 + M_z^2$ (which commute among themselves).

{\it Exercise 1.1} (a) Use the form
\begin{equation}
\label{Mpm}
[M_z, M_{\pm}] = \pm \hbar M_{\pm}\,, \ \ [M_+, M_-] = 2\hbar M_z \ \ {\rm for} \ \ M_\pm = M_x \pm i M_y
\end{equation}
of the CR (\ref{M}) to prove that the spectrum of $M_z$ in any irreducible (finite dimensional) representation of
$SU(2)$ has the form
\begin{equation}
\label{Mz}
(M_z - m\hbar) |j, m> = 0\,, \ \ m = -j, 1-j, \dots ,j-1, j\,,\ \ \ j = 0, \frac{1}{2}, 1, \frac{3}{2}, \dots \ .
\end{equation}
(b) Use the relation
\begin{equation}
\label{M2}
{\bf M}^2 = M_z^2 +\frac{1}{2}(M_+ M_- + M_- M_+) = M_z^2 + \hbar M_z + M_+M_-
\end{equation}
to prove that $({\bf M}^2 - j(j+1)\hbar^2)|j, m> = 0$. 

({\it Hint}: use the relation $M_- |j, -j> = 0 (= M_+|j, j>)$.)

In general, however, there is no ``optimal algorithm'' to quantize a given
classical system. That's why it is often said that {\it quantization is an art}\footnote{Words Ludwig Faddeev used in the discussion after Witten's talk on \cite{GW} in Lausanne, March, 2009, alluding to the Lax ordering in the quantization of integrable systems \cite{F}. A year later Witten's student used the same words as a title of Sect. 2 of \cite{G10}.} Here are three examples in which {\it we know}
what quantization means. The most familiar one is $M = {\mathbb R}^{2n}$
equipped with the {\it canonical symplectic form}
\begin{equation}
\label{dpdq}
\omega = dp\wedge dq = \sum_{j=1}^n dp_j\wedge dq^j,
\end{equation}
with a given choice of affine structure in which $p_i$ and $q^j$ are linear
functions on $M$. Another important example is a cotangent bundle $M = 
T^*{\mathcal Q}$, equipped with a {\it contact form} $\theta = p dq, \, 
q\in{\mathcal Q}, \, p\in T^*_q {\mathcal Q}$, which can be quantized in a natural
 way in terms of a {\it half-density} on ${\mathcal Q}$. Similarly, there is a 
natural procedure to quantize a {\it K\"ahler \footnote{The German mathematician Erich
 K\"ahler (1906-2000) introduced his hermitean metric in 1932 while at the 
University of Hamburg. See about his work and personality R. Brendt, O. 
Riemenschneider (eds), {\it E. K\"ahler, Mathematical Works}, de Gruyter, Berlin
2003; see in particular the articles by S.-S. Chern and by R. Brendt and A.
Bohm. The quantization of K\"ahler manifolds is a lively subject of
continuing interest - see, e.g. \cite{AdPW}, \cite{Hi}, \cite{GW}, \cite{W10},
\cite{G10}. We shall survey it in Sect. 3, below.} manifold} (see Sect. 2.1) 
by taking holomorphic sections of the appropriate line 
bundle. These examples partly overlap. We may, for instance, introduce a 
complex structure on ${\mathbb R}^{2n}$ setting
\begin{equation}
\label{zpq}
\sqrt{2} z_j = q_j - i p_j \, \Rightarrow dp\wedge dq = i dz\wedge d{\bar z}.
\end{equation}
These two ways of viewing the classical phase space are not exactly equivalent,
 however. Each new structure reduces the natural invariance group of the theory.
 If the group preserving the affine structure of ${\mathbb R}^{2n}$ is $GL(2n,
{\mathbb R})$, the symmetry group of the K\"ahler form (\ref{zpq}) is its
subgroup $U(n)$ - the intersection of the orthogonal and the real symplectic 
subgroups of $GL(2n, {\mathbb R}): \, U(n)\simeq O(2n)\cap Sp(2n,{\mathbb R})$.

\bigskip

\section{Introduction to geometric quantization}
\setcounter{equation}{0}
\renewcommand\theequation{\thesection.\arabic{equation}}

\smallskip

 We begin with Baez's explanation \cite{B06} why quantization is a mystery.

``Mathematically, if quantization were 'natural' it would have been a {\it
functor} from the category $Symp$ whose objects are symplectic manifolds
(=phase spaces) and whose morphisms are symplectic maps (=canonical 
transformations)
to the category $Hilb$ whose objects are Hilbert spaces and whose morphisms are
unitary operators.'' Actually, there is a functor from $Symp$ to $Hilb$ which 
assigns to each ($2n$-dimensional) symplectic manifold $M$ (or $(M, \omega)$) 
the Hilbert space $L^2(M)$ (with respect to the measure associated with the symplectic form
 $\omega$ on $M$, given by (\ref{dpdq}) in the simplest case of an affine 
phase space). This is the so called {\it prequantization} which will be 
sketched in Sect. 2.2 below.\footnote{For a reader's friendly review of various 
quantization methods (and a bibliography of 266 titles) - see \cite{AE}. 
For later more advanced texts on prequantization -
 see \cite{WZ, ZZ}; prequantization is the first step to the {\it
geometric quantization} \cite{Wo} of Kostant and Souriau that grew out of
Kirillov's orbit method \cite{K99}. It is reviewed in the very helpful lecture
notes \cite{B}, available electronically, and is the subject of recent research
\cite{H90, AdPW}; it is also briefly discussed among other modern 
approaches to quantization in \cite{GW, G10}.}

\smallskip

\subsection{Elements of symplectic geometry}
\begin{flushright}
{\it Hamiltonian Mechanics is geometry in phase space.}

 Vladimir Arnold (1937-2010), 1978.
\end{flushright}

{\footnotesize {\bf The language of categories.} The reader should not be scared of terms like 
category and functor: this language, introduced by Eilenberg and MacLane and developed by
Grothendieck and his school, has become quite common in modern mathematics and appears to be 
natural for an increasing number of problems in mathematical physics - including quantization 
(and homological mirror symmetry - cf. \cite{G10}). For a friendly introduction  - see \cite{B06}; 
among more advanced mathematical texts the introductory material of \cite{GM}, including the first 
two chapters, is helpful. We recall, for reader's convenience, a couple of
informal definitions. A {\it category} $\mathcal{C}$ consists of a
{\it class of objects}, $X, Y \in \mathcal C$ and of non-intersecting sets of
maps $Hom(X, Y)$, called {\it morphisms} and denoted as $\varphi: X\rightarrow
Y$ whose composition is associative. We note that the definition of a category
only involves operations on morphisms, not on objects. A {\it functor} $F:
\mathcal C \rightarrow \mathcal D$ between two categories is a map
$X\rightarrow F(X)$ between objects, together with a map $\varphi \rightarrow
F(\varphi)$ between morphisms, such that $F(\varphi \psi) = F(\varphi) F(\psi)$
 whenever $\varphi \psi$ is defined; in particular, $F(id_X) = id_{F(X)}$. An
important example is the {\it fundamental group} which may be viewed as a
functor from the category of topological spaces to
the category of groups (with the corresponding homomorphims as morphisms).}

We proceed to defining some basic notions of symplectic differential geometry, 
a subject of continuing relevance for mathematical physics, with a wealth of 
competing texts - see, for instance, \cite{Br, CdS, deG, F, M, V}.

The {\it tangent bundle} $TM$ of a differentiable manifold $M$ is spanned by
{\it vector fields} or {\it directional derivatives}, - i.e., first order
homogeneous differential operators $X^i(x)\partial_i$ that are linear
combinations of the derivatives $\partial_i:= \frac{\partial}{\partial x^i}$ in
 the neighbourhood of each point with local coordinates $x^i$. The cotangent
bundle consists of 1-forms, spanned by the differentials $dx^i$ viewed as
linear functionals on vector fields, such that
\begin{equation}
\label{dx}
dx^i(\partial_j) = (dx^i, \partial_j) = \delta^i_j \, \, (\delta^i_j =
diag(1, ..., 1)), \, {\rm for} \, \partial_j = \frac{\partial}{\partial x^j}.
\end{equation}
One also assumes that $\partial_i$ anticommutes with $dx^j$. Denoting the
contraction with a vector field $X$ by ${\hat X}$ we shall have, for instance,
\begin{equation}
{\hat \partial}_q dp\wedge dq = - dp.
\end{equation}
We note that a contraction of a vector field $X$ with a differential form
$\omega$ is more often written as $i_X \omega$. One also uses the notion of a
tensor field $T^r_s(x)$ (contravariant of rank $r$ and covariant of rank $s$)
defined as an element of the tensor product $(T_x M)^r\otimes(T^*_x M)^s$
(smoothly depending on $x$). In constructing higher rank exterior differential
forms we use the anticommutativity of $d$ with odd differentials; if $r$ is the
rank of the form $\omega_r$, then:
\begin{equation}
\label{d^2}
d(\omega_r\wedge\alpha) = (d\omega_r)\wedge\alpha + (-1)^r\omega_r \wedge d\alpha \, \, (d^2 = 0).
\end{equation}
We say that $\omega_r$ is a {\it closed form} if $d\omega_r = 0$; it is called
{\it exact} if there exist an $(r-1)$-form $\theta$ such that $\omega_r = d\theta$.
Denoting the additive group of closed r-forms by ${\mathcal C}_r$ and its subgroup of
exact forms ({\it boundaries}) by ${\mathcal B}_r$ we define the r-th {\it cohomology
group} as the quotient group
\begin{equation}
\label{Hr}
H^r(M) (=H^r(M, {\mathbb R})) = {\mathcal C}_r /{\mathcal B}_r.
\end{equation}

Another important concept, the {\it Lie\footnote{The Norvegian mathematician Sophus 
Lie (1842-1899) devoted his life to the theory of continuous transformation groups.}
 derivative} ${\mathcal L}_X$ along a
vector field $X$ (for a historical survey - see \cite{Tr}), can be defined
algebraically demanding that: (1) it coincides with the directional derivative
along $X$ on smooth functions: ${\mathcal L}_Xf = Xf$;
(2) it acts as a derivation (i.e., obeys the {\it Leibniz rule}) on products of
 tensor fields:
\begin{equation}
{\mathcal L}_X S\otimes T = ({\mathcal L}_X S)\otimes T +
S\otimes{\mathcal L}_X T;
\end{equation}
(3) it acts by commutation on vector fields: ${\mathcal L}_X Y = [X, Y]$;
(4) acting on a differential form it satisfies {\it Cartan's\footnote{\'Elie 
Cartan (1869-1951) introduced the general notion of antisymmetric differential 
forms (1894-1904) and the theory of spinors (1913); he completed in his 
doctoral thesis (1894) Killing's classification of semisimple Lie algebras.}  
magic formula}
\begin{equation}
{\mathcal L}_X\omega = \hat{X}d\omega + d\hat{X}\omega, \, {\rm in \,
particular}, \, {\mathcal L}_X d\omega = d{\mathcal L}_X \omega.
\end{equation}

A {\it symplectic manifold}: is defined as a manifold with a non-degenerate
closed 2-form. (A non-degenerate 2-form $\omega$ on a $2n$-dimensional manifold is characterized by the fact that 
the corresponding Liouville\footnote{Joseph Liouville (1809-1882) proved that a Hamiltonian time evolution
is measure preserving. His contributions to complex analysis and to number theory are also famous.} volume form 
$\omega^{\wedge n}$ is nonzero.) If one writes the symplectic form in local coordinates as 
$\omega = \frac{1}{2}\omega_{ij}dx^i\wedge dx^j, \, \omega_{ij} = -\omega_{ji}$ then the skew-symmetric 
matrix $(\omega_{ij})$ is invertible and its inverse, $({\mathcal P}^{ij})$, defines {\it Poisson brackets} among functions on $M$:
\begin{equation}
\label{Omega}
\{f, g\} = {\mathcal P}^{ij}\partial_i f \partial_j g \, \, {\rm for} \, \partial_i = \frac{\partial}{\partial x^i}, \, 
 {\mathcal P}^{ik}\omega_{kj} = \delta^i_j.
\end{equation}
Each symplectic manifold is even dimensional and orientable. In the neighbourhood of each point it admits local 
{\it Darboux\footnote{Jean-Gaston Darboux (1842-1917) established the existence of canonical variables in his study of the
 Pfaff problem in 1882.} coordinates} $(p_i, q^j)$ in which the symplectic form $\omega$ is given by the canonical 
expression (\ref{dpdq}). 

To each function $f$ on the symplectic manifold $(M, \omega)$
there corresponds a {\it Hamiltonian\footnote{William Rowan Hamilton (1805-1865)
introduced during 1827-1835 what is now called {\it Hamiltonian} but also the
{\it Lagangian} formalism unifying mechanics and (geometric) optics. He invented 
the quaternions (discussed in Sect. 3.3 below) in 1843.} vector field} 
$X_f$ such that $\hat{X}_f\omega := \omega(X_f, .) = df(.)$. In the above affine
 case it is given by:
\begin{equation}
X_f = \frac{\partial f}{\partial q} \, \frac{\partial}{\partial p} -
\frac{\partial f}{\partial p} \, \frac{\partial}{\partial q} \, .
\end{equation}
The Poisson bracket between two functions on M is expressed in terms of the
corresponding vector fields as
\begin{equation}
\label{PB}
\{f , g\}\ = X_f g = -X_g f = \omega(X_g, X_f) = \frac{\partial f}{\partial q} \, \frac{\partial g}{\partial p} - \frac{\partial f}{\partial p} \, \frac{\partial g}{\partial q} \, .
\end{equation}
It follows that the commutator algebra of vector fields provides a representation
of the infinite dimensional Lie algebra of Poisson brackets:
\begin{equation}
 [X_f, X_g] = X_{\{f , g\}}.
\end{equation}
A smooth manifold $M$ equipped with a Poisson bracket (that is skew-symmetric and satisfies the {\it Jacobi identity} is called
a {\it Poisson manifold}. It is clear from (\ref{PB}) that the PB gives rise to a derivation on the algebra of smooth functions 
on $M$ which obeys the {\it Leibniz\footnote{Gottfried Wilhelm Leibniz (1646-1716), mathematician-philosopher, a precursor of the symbolic logic, codiscoverer of the calculus - together with Isaac Newton (1642-1727).} rule}, thus defining a {\it Poisson structure}:
\begin{equation}
\label{Leib}
\{f, gh\} = \{f, g\} h + g\{f, h\}. 
\end{equation}
Poisson manifolds (which are symplectic if and only if the matrix ${\mathcal P}$ that defines a {\it Poisson bivector} is
invertible) are the natural playground of deformaton quantization (surveyed in Sect. 4 below).
 
A {\it compact symplectic manifold} should necessarily has a nontrivial second cohomology 
group. It follows that a sphere ${\mathbb S}^n$ only admits a symplectic 
structure for $n = 2$. 

{\it Exercise 2.1} Demonstrate that the 1-form
\begin{equation}
\label{H1}
\eta_1 = i\frac{zd{\bar z} - {\bar z}dz}{2z{\bar z}} = \frac{xdy - ydx}{x^2 + 
y^2}
\end{equation}
in the punctured plane ${\mathbb C}^* = \{z=x+iy, \, (x,y)\in{\mathbb R}^2; \,
r^2:= x^2 + y^2 > 0\}$ is closed but not exact, albeit locally, around any non-zero
point $(x, y)$, it can be written as a differential of a multivalued function,
\begin{equation}
\label{dphi}
\eta_1 = d\varphi \, {\rm for} \, \varphi = arcsin\frac{y}{r}
= arccos\frac{x}{r} = arctg\frac{y}{x}.
\end{equation}
Prove that if $\eta$ is an arbitrary element of $H^1(C^*)$, i.e. if $\int_{{\mathbb S}^1}
\eta = b\neq 0$ then the 1-form $\eta - \frac{b}{2\pi} \eta_1$ is exact.
({\it Hint}: use the fact that the integral of $d\varphi$ along the unit circle is $2\pi$.)

A {\it (pseudo)Riemannian\footnote{The great German mathematician Bernhard Riemann (1826-1866) 
introduced what we now call Riemannian geometry in his inaugural (in fact test) lecture 
in 1854. More about Riemann and his work can be found in \cite{Mo}.} manifold} is a real
differentiable manifold $M$ equipped with a nondegenerate quadratic form $g$ at each point 
$x$ of the tangent space $TM$ that varies smoothly from point to point. We shall be mostly 
interested in the case of {\it Riemannian metric} in which the form $g$ is {\it positive 
definite}. An n-dimensional {\it complex manifold} can be viewed as a 2n dimensional real
manifold equipped with an integrable {\it complex structure} - i.e., a vector bundle  
endomorphism $J$ of $TM$ (that is a tensor field of type $(1, 1)$) such that $J^2 = -1$.

{\footnotesize Such an endomorphism (i.e. a linear map of $TM$ to itself) of square $-1$
is called an {\it almost complex structure}. An almost complex structure $J$ and a Riemannian 
metric $g$ define a {\it hermitean\footnote{Named after the French mathemtician
 Charles Hermite (1822-1901), the first to prove that the base $e$ of natural 
logarithms is a transcendental number.} structure} if they satisfy the 
compatibility condition
\begin{equation}
\label{gJ}
g(JX, JY) = g(X, Y).
\end{equation}
Every almost hermitean manifold admits a nondegenerate {\it fundamental 2-form}
 $\omega (= \omega_{g,J})$:
\begin{eqnarray}
\label{omgJ}
\omega(X, Y):= g(X, JY) \, \Rightarrow \omega(X, Y) = \omega(JX, JY) \nonumber \\ 
= g(JX, J^2Y) =- g(JX, Y) = -g(Y, JX) = -\omega(Y, X).  
\end{eqnarray}
If the almost complex structure is covariantly constant with respect to the Levi-Civita\footnote{The 
Italian mathematician Tullio Levi-Civita (1873-1941) is known for his work on {\it absolute 
differential (tensor) calculus.}} connection then the fundamental form is closed (and hence symplectic):
\begin{equation}
\label{nabla}
\bigtriangledown J = 0 \, \Rightarrow d\omega_{g,J}=0
\end{equation}
(and, moreover, the so called Nijenhuis tensor $N_J$ (of rank (2, 1)), related to $J$, vanishes;
this provides an integrability condition which is necessary and sufficient for the almost
complex structure to be a {\it complex structure}).} 

The endomorphism $J$ of the tangent bundle $TM$ defines an {\it integrable} complex 
structure if $M$ is a complex manifold with a holomorphic atlas (including holomorphic transition 
functions) on which the operator $J$ acts as a multiplication by $i$. A {\it K\"ahler manifold} is
a Riemannian manifold with a compatible complex structure. (Introductory 
lectures on complex manifolds in the context of Riemannian geometry are 
available in \cite{V} .  For a more systematic study of K\"ahler manifolds the 
reader may consult the lecture notes \cite{Ba} and \cite{M}. We shall deal with
 the quantization of ${\mathbb C}^n$ as a K\"ahler manifold in Sect. 3.)

Complex forms admit a unique decomposition into a sum of $(p, q)$-forms that are
 homogeneous of degree $p$ in $dz^i$ and of degree $q$ in $dz^j$. The 
differential $d$ can be decomposed into {\it Dolbeault differentials} 
$\partial$ and ${\bar \partial}$ which increase $p$ and $q$, respectively:
\begin{equation}
\label{Dolb} 
\partial = dz\wedge\frac{\partial}{\partial z}, \, {\bar \partial} = 
d{\bar z}\wedge\frac{\partial}{\partial{\bar z}} \ \ (d = \partial +
 {\bar \partial}).
\end{equation}
Similarly, one defines the {\it Dolbeault cohomology groups} $H^{p,q}$.

\smallskip

\subsection{Prequantization}

We shall see that even this first, better understood step to quantization does
not always exist: it imposes some restrictions on the classical mechanical data; on
the other hand, it requires the addition of some extra structure (a comlex line
bundle) to it, whose properties may vary. In other words, when prequantiztion is
possible, it is not, in general, unique.

The functions on $M$ play two distinct roles in the prequantization: first, the real
smooth functions $f(p, q)$ span the Poisson algebra $\mathcal{A}$ of (classical) observables;
second, the ``prequantum states'' are vectors (complex functions $\Psi(p, q)$ on $M$,
square integrable with respect to the Liouville measure) in a Hilbert space $\mathcal{H}$.
The prequantization requires to equip $M$ with a {\it complex line bundle $L$}. Another fancy
way to state this is to say that the {\it wave function} (both quantum and "prequantum") is
a $U(1)$-{\it torsor} - only relative phases (belonging to $U(1)$) have a physical meaning.
(For an elementary, physicist-oriented, introduction to the notion of torsor - see
\cite{B09}.)  We are looking for a {\it prequantization map} $\mathcal{P}: \mathcal{A} \rightarrow
\mathcal{P}\mathcal{A}$ where $\mathcal{P}\mathcal{A}$ is an operator algebra of ``prequantum observables''
acting on $\mathcal{H}$ and satisfying:

(i) ${\mathcal P}(f)$ is linear in $f$ and ${\mathcal P}(1) = {\bf 1}$
(the identity operator in ${\mathcal H}$);

(ii) it maps the Lie algebra of Poisson brackets into a commutator algebra:
\begin{equation}
\label{PPB}
 [{\mathcal P}(f),{\mathcal P}(g)] = i \hbar {\mathcal P}(\{f, g\})\,.
\end{equation}
(One may also assume a functoriality property - covariance under mapping of one symplectic manifold
to another - see e.g. requirement (Q4) in Sect. 3 of \cite{AE}.) The vector fields ${\mathcal P}(f)= i \hbar
X_f$ obey (\ref{PPB}) but violate condition (i) (since $X_1 = 0$). There is, however, a (unique) inhomogeneous
first order differential operator which does satisfy both properties for the affine phase space:
\begin{equation}
 \label{P(f)}
{\mathcal P}(f) = i \hbar X_f + f + \theta(X_f)\,, \quad \theta = p dq \quad (\omega = d\theta)\,.
\end{equation}

{\it Exercise 2.2} Verify (using $\theta(X_f) = 
-p\frac{\partial f}{\partial p})$ that
\begin{equation}
[{\mathcal P}(f), {\mathcal P}(g)] = {\mathcal P}(\{f, g\}).
\end{equation}

If we identify $f$ with the classical Hamiltonian $H$ then the term added to
$X_H$ is nothing but (minus) the Lagrangian: $H - p\partial{H}/\partial{p} =
-\mathcal{L}$. Viewing $H$ as the generator of time evolution and integrating in
 time we see that the resulting phase factor in the wave function is highly
reminiscent to the Feynman path integral.

For the coordinate and momentum Eq. (\ref{P(f)}) gives, in particular,
\begin{equation}
\label{Pqp}
{\mathcal P}(q) = q + i\hbar \partial/\partial{p}\,, \ \ {\mathcal P}(p) =
-i \hbar \partial/\partial{q}\,.
\end{equation}
We observe that our prescription sends real observables $f$ to hermitean
operators\footnote{One actually needs {\it selfadjoint operators} in order
to ensure reality of their spectrum but we won't treat here the subtleties
with domains of the resulting unbounded operators.} (a requirement hidden
in the correspondence with (\ref{qp})):

(iii) ${\mathcal P}(f)^* = {\mathcal P}(f)$ for real smooth functions $f(p, q)$.

The association $f \rightarrow {\mathcal P}(f)$ is,
nevertheless, physically unsatisfactory since it violates simple algebraic
relations between observables. For instance, the prequantized image of the
kinetic energy of a nonrelativistic particle,
\begin{equation}
\label{H0}
H_0 = \frac{{\bf p}^2}{2m}, \, \, {\bf p}^2 = p_1^2 + p_2^2 + p_3^2,
\end{equation}
is
\begin{equation}
\label{PH0}
{\mathcal P}(H_0) = -i\hbar {\bf \partial}_{\bf p} H_0 {\bf \partial}_{\bf q} -
H_0 \neq H_0({\mathcal P}({\bf p})).
\end{equation}
The operator ${\mathcal P}(H_0)$ violates, in particular, energy positivity.

The definition (\ref{P(f)}) applies whenever $M$ is a {\it cotangent bundle}, 
$M = T^*\mathcal{Q}$, so that the symplectic form is exact, $\omega = d\theta$.
 This is never the case for a compact phase manifold (that would have had 
otherwise a zero volume). In general, prequantization requires that 
$\omega/{2\pi\hbar}$ represents an integral cohomology class in 
$H^2(M,\mathbb{R})$ - i.e., that its integral over any closed (orientable) 
2-surface in $M$ is an integer. These are, essentially, the Bohr-Sommerfeld(-
Wilson)\footnote{Bohr's model was further developped by Arnold Sommerfeld 
(1868-1951). Four among his doctoral students in Munich won the Nobel Prize in 
Physics. Sommerfeld himself was nominated for the prize 81 times, more than any
 other physicist. The British physicist William Wilson (1875-1965) discovered 
independently the quantization conditions in 1915.} quantization conditions, 
discovered in 1915, before the creation of quantum mechanics. For instance, a 
2-sphere of radius $r, {\mathbb S}_r^2$ is (pre)quantizable (for a fixed value 
of the Planck constant $\hbar$) iff $r = n\hbar/2, n\in \mathbb Z$. In either 
case, the symplectic form does not change if we add  an exact form $df$ to the 
contact form  $\theta$, satisfying (locally or globally) $d\theta = \omega$. 
Such a change can be compensated by multiplying the elements of our Hilbert 
space ${\mathcal L}^2(M, \omega)$ by the phase factor $exp(if/\hbar)$. This 
suggests that it is more natural to regard $P(f)$ as acting on the space of 
sections of a complex line bundle $L$ over$M$ equipped with a connection $D$ of 
the form:
\begin{eqnarray}
\label{D}
D = d-\frac{i}{\hbar} \theta\,, \ \ d = dx^i \partial_i\,, \ \ \partial_i 
\equiv \frac{\partial}{\partial{x^i}}\ \ \Rightarrow \nonumber \\
D_X = X - \frac{i}{\hbar} \theta(X) \ \ (X=X^i\partial_i\,,\ \theta=\theta_i 
dx^i\ \Rightarrow\ \theta(X)=\theta_i X^i)\,,
\end{eqnarray}
where $X$ is an arbitrary (not necessarily Hamiltonian) vector field, $x^i$ are
 local coordinates on $M$ (and we use the summation convention for repeated 
indices). The curvature form of this connection coincides with our symplectic 
form $\omega$:
\begin{eqnarray}
\label{RD}
&&R(X,Y):=i([D_X, D_Y]-D_{[X,Y]})=\frac{1}{\hbar}(X\theta(Y)-Y\theta(X)-
\theta([X, Y])) \nonumber \\ &&=\frac{1}{\hbar} d\theta(X,Y)= \frac{1}{\hbar} 
\omega(X,Y).
\end{eqnarray}

{\footnotesize In order to get an idea how an integrality condition arises from the existence of a hermitean connection compatible with the symplectic structure on a general phase manifold we should think of an {\it atlas of open neighbourhoods} $U_\alpha$ covering the manifold $M$.
The quantum mechanical wave function is substituted by a section of our complex line bundle. It is given by a complex valued function $\Phi_\alpha$ on each 
chart $U_\alpha$ and a system of transition functions $g_{\alpha \beta}$ for each non-empty intersection $U_{\alpha\beta} = U_\alpha\cap U_\beta$, such that 
$\Phi_\alpha = g_{\alpha\beta} \Phi_\beta$ on $U_{\alpha \beta}$. Consistency for double and triple intersections requires the {\it cocycle condition}:
\begin{equation}
\label{cocycle}
g_{\alpha \beta} g_{\beta \alpha} = 1, \, \, g_{\alpha \beta} g_{\beta \gamma} g_{\gamma \alpha} = 1. 
\end{equation}
If the contact 1-forms $\theta_\alpha$ are related in the intersection $U_{\alpha \beta}$ of two 
charts by $\theta_{\alpha} = \theta_\beta + du_{\alpha \beta}$ then the hermiticity of the 
connection and the cocycle condition imply integrality of the (additive) cocycle of $u_{\alpha \beta}$:
\begin{equation}
\label{g-u}
g_{\alpha \beta} = exp(i\frac{u_{\alpha \beta}}{\hbar}) \, \Rightarrow 
u_{\alpha \beta} + u_{\beta \gamma}      + u_{\gamma \alpha} = h n_{\alpha \beta \gamma}
 \, \, {\rm where} \, \, n_{\alpha\beta\gamma} \in {\mathbb Z}.
\end{equation}
The theorem that the above stated integrality condition for the symplectic form is necessary and sufficient
for the existence of a hermitian line bundle $L$ with a compatible connection $D$ whose curvature is
$\omega$ goes back (at least) to the 1958 book of Andr\'e Weil (1906-1998) \cite{W}.)} 

Looking at the example of the 2-sphere one can get the wrong impression that the integrality condition 
for $\omega \in H^2(M,\mathbb{R})$ can be always satisfied by just rescaling the symplectic form. The 
simple example of the product of two spheres ${\mathbb S}_r^2\times{\mathbb S}_s^2$ with incommensurate 
radii (i.e. for irrational $r/s$) shows that this is not the case: there are (compact) symplectic 
manifolds that are not prequantizable.

The equivalence classes of prequantizations (whenever they exist) are given by the first cohomology group
of $M$ with values in the circle group $U(1)$ or equivalently by the ($U(1)$-valued) characters of the
fundametal group of $M$:
\begin{equation}
\label{H1U1}
H^1(M, U(1)) = \pi_1(M)^*.
\end{equation}
We shall illustrate this statement on the example of the cotangent bundle to 
the circle, that is, on the cylindric phase space
\begin{equation}
\label{cyl}
M = T^*{\mathbb S}^1\,,\ \ \omega = dp\wedge d\varphi = d \theta\,, \ \
\theta = p\, d\varphi\,,\ \ p \in {\mathbb R}\,, \ \ 2 \pi\varphi \in 
\mathbb{R}/\mathbb{Z}.
\end{equation}
The fundamental group of the circle being $\pi_1({\mathbb S}^1) = \mathbb{Z}$ 
the group of its characters coincides with $U(1)$. We thus expect to have a 
continuum of inequivalent prequantizations of $(M, \omega)$
labeled by elements of $U(1)$. This can be realized by adding to the connection
 $D$ the closed form $i\lambda\,
d\varphi$   ($d\varphi$ is not exact since $\varphi$ is not a global coordinate on the circle). Inserting
this in (\ref{Pqp}) we find ${\mathcal P}_\lambda (p) = \hbar \lambda - i\hbar 
\frac{\partial}{\partial \varphi}$ which  gives rise to $\lambda$-dependent 
inequivalent prequantizations for $\lambda \in [0, 1)$.

\smallskip

\subsection{From prequantization to quantization}

\hfill\begin{minipage}{.7\linewidth}
{\it Now it doesn't seem to be true that God created a classical universe on the first day
and then quantized it on the second day.} 
John Baez \cite{B06}
\end{minipage}

\smallskip

In spite of the necessary restrictions for its existence and of its non-uniqueness, prequantization
appears to provide a nice map from a sufficiently wide class of complex line bundles over classical
phase spaces to naturally defined operator algebras on Hilbert spaces, so that our conditions (i), (ii)
(involving Eq. (\ref{PPB})) and (iii) are indeed satisfied. This procedure does have a shortcoming of
excess, however: the resulting prequantized algebra and the corresponding Hilbert space are much too big.
Matthias Blau, \cite{B}, includes in his list of desiderata the following {\it irreducibility requirement}.
Consider a {\it complete set of classical observables}, like $p_i$ and $q^j$ in the simplest case of an
affine phase space, such that every classical observable is a function of them; alternatively, we can
characterize a complete set $(f_1, ..., f_n)$ by the property that the only classical observables
which have zero Poisson brackets with all of them are the constants. Blau then demands that their images
$(Q(f_1), ..., Q(f_n))$  under the {\it quantization map} $Q(f)$ (from the algebra $\mathcal{A}$ of
classical observables to the quantum operator algebra $\mathcal{A}_\hbar$) are {\it operator
irreducible}, that is if an operator $A$ in  $\mathcal{A}_{\hbar}$ commutes with all $Q(f_j)$
then it should be a multiple of the identity. If we allow all operators in Hilbert space ${\mathcal L}^2(M, \omega)$
then we see that the prequantization violates this condition: the operator $p -{\mathcal P}(p) = p + i\hbar
\frac{\partial}{\partial{q}}$ commutes with all ${\mathcal P}(p),{\mathcal P}(q)$ (without being a multiple
of the identity). One may disagree with this objection on the ground that multiplication operator by  $p$
is not of the form ${\mathcal P}(f)$. The physical shortcoming, indicated in Sect. 2.2: the fact that the
prequantized nonrelativistic kinetic energy (\ref{PH0}) is not proportional to the square of the prequantized
momentum and is not a positive operator appears to be more serious. We shall therefore look for a quantization
map $Q$ which satisfies - along with the conditions (i), (iii) (and a weakened version of (ii)) - a condition 
that would guarantee the positivity of the quantum counterpart of the square of a real observable. The following 
requirement appears to achieve this goal in a straightforward manner.

(iv) If $Q(f)$ is the image of the real observable $f$, then one should have $Q(f^2) = Q(f)^2$.

{\footnotesize {\it Remark 2.1} In the framework of (formal) deformation quantization - see Sect. 4.2 - one can only 
assume such an equality up to terms of order $\hbar^2$. According to the Darboux theorem (generalized by Sophus Lie (1842-1899)
- see Sect. 4.2), every symplectic manifold admits canonical coordinates with a locally constant Poisson bivector. A weaker  
requirement that would be sufficient to ensure the positivity of the kinetic energy on a cotangent bundle, consists in just 
demanding the validity of (iv) for functions of the canonical momenta.} 

Baez \cite{B06} conjectures that there is no positivity preserving functor from the symplectic category to
the Hilbert category. In fact, there is a result of this type (of Groenewold and van Hove)\footnote{We shall say more about the Dutch
theoretical physicist H.J. Groenewold and about his paper \cite{G46} in Sect. 4.1 below. In a pair of 1951 papers
the Belgian physicist Leon van Hove (1924-1990) refined and extended Groenewold's result, showing effectively
that there exists no quantization functor consistent with Schr\"odinger's quantization of ${\mathbb R}^{2n}$.}
for the algebra of polynomials of $p$ and $q$ in an  affine phase space. One has to 
settle to a weaker version of requirement (ii) only demanding the validity of 
(\ref{PPB}) (with $\mathcal{P}$ replaced by $Q$) for some ``suitably chosen'' 
Poisson subalgebra of the algebra of observables. Quantization becomes an art 
for the physicist and a mystery for the mathematician. To give a glimpse of 
what else is involved in the geometric quantization we shall sketch 
the next step in the theory, defining the notion of a polarization.

The quest for a mathematical understanding started {\it after} the art of quantization was mastered and
displayed on examples of physical interest. Rather than following a mathematical intuition, geometric quantization
attempts to extract general properties of such known examples. The first observation is that the state vectors
should only depend on half of the phase space variables, like in the Schr\"odinger picture. More precisely, one
should work with wave functions depending on a {\it maximal set of Poisson commuting observables}.  The right
way to eliminate half of the arguments is to consider sections of our line bundle that are covariantly constant
along an n-dimensional ``integrable'' subbundle $S$ of vector fields. In other words, our wave functions $\Psi$
should satisfy a system of {\it compatible equations}:
\begin{equation}
\label{CCon}
D_X\Psi = 0\,, \ \ X\in S \ \ \Rightarrow [D_X, D_Y]\Psi = 0 \ \ {\rm for} \ \ X, Y \in S\,.
\end{equation}
It is clear from (\ref{RD}) that if the subbundle $S$ is closed under commutation (in other words, if
$X, Y \in S \Rightarrow [X, Y] \in S$, that is, if the vector fields in $S$ are in involution) and if in
addition the corresponding integral manifold is {\it (maximally) isotropic} - i.e., $\omega(X, Y) = 0$ for $X, Y \in S$
(and $dim S = \frac{1}{2} dim M = n$), then the compatibility (also called {\it integrability}) condition in (\ref{CCon}) 
is automatically satisfied.  Such maximally isotropic submanifolds are called {\it Lagrangian}. We are tacitly
assuming here that the dimensionality of $S$ does not change from point to point. This is not an innocent assumption.
For $M = {\mathbb S}^2$ it means that a polarization would be given by a nowhere vanishing vector field. On the other
hand, it is known that there is no such globally defined vector field on the 2-sphere. (In fact, among the closed
2-dimensional surfaces only the torus has one.) The way around this difficulty is to complexify the (tangent bundle of
the) phase space. Integrable Lagrangian subbundles are indeed more likely to exist on $TM_{\mathbb{C}}$ than on $TM$.
Thus we end up with the following {\bf definition}. {\it A polarization of a symplectic manifold $(M, \omega)$ is an
integrable maximal isotropic (Lagrangian) subbundle $S$ of the complexified tangent bundle $TM_{\mathbb{C}}$ of $M$.}

We shall consider examples of two opposite types: real, $S=\bar{S}$, and 
K\"ahler polarizations, $S \cap \bar{S} = \{ 0 \}$, the types most often 
encountered in applications. As real polarizations are the standard lore of 
elementary quantum mechanics we shall mention them only briefly, while 
devoting a separate section to (complex) K\"ahler polarizations.

A {\it real polarization} is encountered typically in a cotangent bundle,
$M = T^*{\mathcal Q}$. In local coordinates $S$ is
spanned by the vertical vector fields $\partial/\partial p$, yielding the
standard Schr\"odinger representation in which
the coordinates are represented as multiplication by $q$ (rather than by the
prequantum operator ${\mathcal P} (q)$ (\ref{Pqp})). When ${\mathcal Q}$ involves a circle
(on which there is no global coordinate) it is advantageous to replace the 
multivalued coordinate $\varphi$ by a periodic function as illustrated on the
simplest example of this type $T^*{\mathbb S}^1$ with contact form $\theta =
pd\varphi$. In this case one can introduce global sections $\Psi$ (satisfying 
$\frac{\partial}{\partial p} \Psi =0$) as analytic functions of $e^{\pm i\varphi}$. Then 
the spectrum of the momentum operator is discrete:
\begin{equation}
\label{pS1}
Q(p) = i\hbar X_p  = -i\frac{\partial}{\partial \varphi} \, \, \Rightarrow
\, (Q(p) - n\hbar) e^{in\varphi} = 0, \, \, n\in{\mathbb Z}.
\end{equation}
There is no symmetry between coordinate and momentum in this example.
 As discussed in \cite{B} the momentum space picture does not always exist 
in $T^*{\mathcal Q}$ and when it does it may involve some subtleties.

The question arises how to define the inner product in the ``physical Hilbert
space'' of polarized sections, - i.e., of functions on ${\mathcal Q}$. We cannot
 use the restriction of the Liouville measure since the integral over the fiber
 diverges (for functions independent of $p$). If ${\mathcal Q}$ is a Riemannian 
manifold, if, for instance, a metric is given implicitly via 
the kinetic energy, we can use the corresponding volume form on it. In general,
however, there is no canonical measure on the quotient space $M/S \sim{\mathcal Q}$. 
The geometric quantization prescribes in this case the use of a {\it half density}, 
defined in terms of the square root of the {\it determinant bundle} $Det{\mathcal Q}
= \Lambda^n T^*{\mathcal Q}$, the n-th skewsymmetric power of the cotangent bundle 
(see \cite{AE}, \cite{B}).

\bigskip

\section{Quantization of K{\"a}hler manifolds}
\setcounter{equation}{0}
\renewcommand\theequation{\thesection.\arabic{equation}}

\smallskip

\subsection{Complex polarization. The Bargmann space}

\smallskip

A (pseudo)K\"ahler manifold can be defined as a complex manifold equipped with a
non-degenerate hermitean form whose real part is a (pseudo)Riemannian metric 
and whose imaginary part is a symplectic form (see Sect. 2.1). Just as the real
 affine symplectic space $({\mathbb R}^{2n}, \omega = dp \wedge dq)$ serves as a
 prototype of a symplectic manifold with a real polarization, the complex space 
${\mathbb C}^n$, equipped with the hermitean form
\begin{equation}
\label{hf}
dz\otimes d\bar{z}\, (\equiv \sum_1^n dz_j\otimes d\bar{z}_j) = g -i\omega\,, \
 \ \omega=idz\wedge d\bar{z}
\end{equation}
($g= \frac{1}{2}(dz\otimes d\bar{z} + d\bar{z}\otimes dz)$), can serve as a
prototype of a K\"ahler manifold. More generally, locally, any (real) K\"ahler 
form can be written (using the notation (\ref{Dolb})) as 
\begin{equation}
\label{Kf}
\omega =i \partial {\bar \partial} K, \, \,  (K = \bar K,  \,  
 d = \partial + {\bar \partial}).
\end{equation}

{\footnotesize It is instructive to start, alternatively, with a real 2n-dimensional symplectic vector space $(V= {\mathbb R}^{2n}, \, \omega)$ . A {\it complex structure} is a (real) map $J: V \rightarrow V$ of square $-1$ - 
see Sect. 2.1. (A 2-dimensional example is provided by the real skewsymmetric matrix $\epsilon:=i\sigma_2$ where $\sigma_j$ are the hermitean Pauli matrices.) Such 
a $J$ gives $V$ the structure of a complex vector space: the multiplication by a complex number $a+ib$ being defined by $(a+ib)v = av +bJv$. The complex structure 
$J$ is compatible with the symplectic form $\omega$ if
\begin{equation}
\label{comp}
\omega(Ju, Jv) = \omega(u, v) \, \, {\rm for \,  all} \, \, u, v \in V.
\end{equation}  
Then $g(u, v):= \omega(Ju, v)$ defines a non-degenerate symmetric bilinear 
form while the form $h(u, v) = g(u, v) - i\omega(u, v)$ is (pseudo)hermitean. We shall 
restrict our attention to {\it K\"ahler} (rather than pseudo-K\"ahler) forms 
for which $g$ and $h$ are positive definite.}

In our case (i.e. for $\omega$ appearing in (\ref{hf})) the K\"ahler potential $K$ and the contact form $\theta$ are given by
\begin{equation}
\label{Ktheta}
K = z \bar z, \, \, \theta = \frac{i}{2}(zd\bar z - \bar z dz).
\end{equation}
The Hamiltonian vector fields corresponding to $z$ and $\bar{z}$ are then:
\begin{equation}
\label{z/bar}
X_z=i\frac{\partial}{\partial{\bar z}}, \, X_{\bar z}=
-i\frac{\partial}{\partial z} \, \, \Rightarrow \{z,\bar z\} = i.
\end{equation} 
We define the complex polarization in which $Q(z) = z$ by introducing sections annihilated by the covariant derivative (\ref{D})
\begin{equation}
\label{Dbar}
\bar D:= D(\frac{\partial}{\partial{\bar z}}) = 
\frac{\partial}{\partial{\bar z}} + \frac{1}{2\hbar}(zd\bar z - \bar z dz) 
(\frac{\partial}{\partial{\bar z}} ) = \frac{\partial}{\partial{\bar z}}  + 
\frac{z}{2\hbar}. 
\end{equation}
The general covariantly constant section, - i.e., the general solution of
the equation  $\bar D \Psi = 0$ is
\begin{equation}
\label{Psi}
\Psi (z, \bar z) = \psi(z) exp(-\frac{K}{2\hbar}), \, \, K=z\bar z,  
\end{equation}
 where $\psi(z)$ is any entire analytic function of $z$ with a finite 
norm square
\begin{equation}
\label{Knorm}
||\Psi||^2 = \int |\psi(z)|^2 exp(-\frac{K}{\hbar})d^{2n}z < \infty \, \,
(d^{2n}z \sim \omega^n). 
\end{equation}
The Hilbert space ${\mathcal B}(= {\mathcal B}_n)$ of such entire functions 
has been introduced and studied by Valentine Bargmann (1908-1989), \cite{B61}, 
and we shall call it {\it Bargmann space}. The multiplication by $z$ plays 
the role of a {\it creation operator} $a^*$. The corresponding annihilation operator has the form 
\begin{equation}
\label{a}
a:=Q(\bar z)= i\hbar X_{\bar z} + \frac{\partial}{2 \partial z} K = 
\hbar\frac{\partial}{\partial z} + \frac{1}{2}\bar z,
\end{equation}
the second term being determined by the condition that $a$ commutes with the covariant derivative $\bar D$.

{\it Exercise 3.1} Prove that $a$ and $a^*$ are hermitean conjugate to each other with respect to the scalar product in ${\mathcal B}$ defined by (\ref{Knorm}) and satisfy the CCR
\begin{equation}
\label{aa*}
[a_i, a_j] =0 = [a^*_i, a^*_j] \,, \ \ [a_i, a^*_j] = \hbar\delta_{ij}.
\end{equation}
Identify ${\mathcal B}_n$ with the Fock space of n creation and n annihilation 
operators with vacuum vector given by (\ref{Psi}) with $\psi(z) = 1$:
\begin{equation}
\label{vac}
\vac = exp(-\frac{K}{2\hbar}), \, \, a_i\vac = 0 = <0|\, a_j^*.
\end{equation} 

{\it Remark 3.1} Recalling the change of variables (\ref{zpq}) we observe that the quantum harmonic oscillator Hamiltonian $H_0$ corresponds to the symmetrized product
of $a^*$ and $a$:
\begin{equation}
\label{osc}
H_0 := \frac{1}{2}(p^2 + q^2) = \frac{1}{2}(a^* a + a a^*) = 
\hbar (z\frac{\partial}{\partial z} + \frac{n}{2}) + \frac{1}{2}z\bar z.
\end{equation}
The additional term $\frac{n}{2}$ coming from the Weyl ordering reflects the fact that the Fock (Bargmann) space carries a representation of the metaplectic group
$Mp(2n)$ (the double cover of $Sp(2n, {\mathbb R})$) \cite{W64} - see also \cite{F}, \cite{deG}, \cite{T10} and references therein. For another treatment of the harmonic 
oscillator, using half forms, see \cite{B}.

\smallskip

\subsection{The Bargmann space ${\mathcal B}_2$ as a model space for $SU(2)$}
 
\smallskip
                                 
We shall now consider the special case $n=2$ of (\ref{hf}), that provides a model of the irreducible representations of $SU(2)$.  This example is remarkably rich. In what follows we shall (1) outline the result of Julian Schwinger (1918-1994) \cite{Sc} and Bargmann \cite{B62} (reproduced in \cite{QTAM}) on the representation theory of $SU(2)$ as a quantization problem and will indicate its generalization to arbitrary semi-simple compact Lie groups; (2) consider the constraint
\begin{equation}
\label{zp}
z\bar{z} (= z_1\bar{z}_1 + z_2\bar{z}_2) = \hbar N   
\end{equation}
where $N$ is any fixed positive integer and study the corresponding gauge theory which gives rise to the quantization of the 2-sphere.
(3) In the next subsection we shall display the {\it hyperk\"ahler structure} (\cite{Hi}) of
 ${\mathbb C}^2$ thus introducing, albeit in a rather trivial context, some basic concepts 
exploited recently - in particular, by Gukov and Witten \cite{GW}, \cite{G10}, \cite{W10}.

To begin with, we note that any Bargmann space ${\mathcal B}$ splits in an (orthogonal) direct sum of
subspaces of homogeneous polynomials $\psi(z) = h_k(z) \, (h_k(\rho z) = \rho^k h_k(z))$. Indeed, the
associated wave functions $\Psi_k$ (\ref{Psi}) span eigensubspaces of $H_0$ of eigenvalues $(k + 
\frac{n}{2})\hbar$, so that polynomials of different degrees k are mutually orthogonal. For $n=2$ the 
eigenvalues $N$ of $H_0/\hbar$ comprise all positive integers and give the dimensions of the corresponding 
eigensubspaces carrying the irreducible representations of $SU(2)$, each appearing with multiplicity one. 

{\footnotesize This construction extends to an arbitrary semi-simple compact Lie group $G$ by considering the subbundle of the cotangent bundle $T^*G$ obtained
 by replacing the fibre at each point by the conjugate
to the Cartan subalgebra of the Lie algebra of $G$ (treated in the case of $G = SU(n)$ and its q-deformation in \cite{HIOPT}).} 

We now proceed to the study of the finite dimensional gauge theory generated by the constraint (\ref{zp}) that gives rise to the eigensubspaces of the 
oscillator's Hamiltonian (\ref{osc}). This Hamiltonian constraint is obviously invariant under $U(1)$ phase transformations generated by its Poisson brackets 
with the basic variables. Using the PB (\ref{z/bar}) and the CCR (\ref{CCR}) and regarding $N$ of Eq. (\ref{zp}) first as a classical and then as a quantum 
dynamical variable we find: 
\begin{equation}
\label{U1}
\{N, z\} = -iz, \, \, e^{iN\alpha}ze^{-iN\alpha} = e^{i\alpha}z, \, e^{iN\alpha}ae^{-iN\alpha} =
  e^{-i\alpha} a.
\end{equation}
For a fixed $N$ Eq. (\ref{zp}) defines a 3-sphere ${\mathbb S}^3$ in ${\mathbb C}^2 \sim {\mathbb R}^4$; it 
can be viewed, according to (\ref{U1}) as a U(1) fibration over the 2-sphere ${\mathbb S}^2(={\mathbb S}^2(\hbar N))$ 
(known as the {\it Hopf fibration}). 

{\footnotesize Heinz Hopf (1894-1971) has introduced this fibration in 1931. It belongs to a family of just three (non-trivial) fibrations in which the total space, the base space, and the fibre are all spheres (and the following sequences of homomorphisms are exact):
\begin{equation}
\label{Hopf}
0 \rightarrow {\mathbb S}^1 \hookrightarrow {\mathbb S}^3 \rightarrow {\mathbb S}^2
\rightarrow 0; \, \, 0 \rightarrow {\mathbb S}^3 \hookrightarrow {\mathbb S}^7 
\rightarrow {\mathbb S}^4 \rightarrow 0;  0 \rightarrow {\mathbb S}^7
\hookrightarrow {\mathbb S}^{15} \rightarrow {\mathbb S}^8 \rightarrow 0.
\end{equation}
This fact is related to the theorem of Adolf Hurwitz (1859-1919) identifying the {\it normed division algebras} 
with {\it the real and the complex numbers, the quaternions and the octonions} - see, e.g., \cite{B02}. (The reals correspond 
to the sequence ${\mathbb S}^0 \hookrightarrow {\mathbb S}^1 \rightarrow {\mathbb S}^1$ where ${\mathbb S}^0 = \{ \pm 1 \}$.)} 

In order to display (and quantize) the symplectic structure of ${\mathbb S}^2(\hbar N))$ it is advantageous to introduce three gauge invariant coordinates obeying one relation (cf. our treatment of ${\mathbb S}^1$ in Sect. 2.3):
\begin{equation}
\label{ksi}
\xi_j = z\sigma_j \bar z, \, j=1, 2, 3, \, \Rightarrow {\bf \xi}^2 = 
(z \bar z)^2 = (\hbar N)^2.  
\end{equation}

The reduction of the form $\omega$ (\ref{hf}) to the 2-sphere (\ref{ksi}) is expressed in terms of the 
Poincar\'e\footnote{Jules Henri Poincar\'e (1854-1912)
introduced his residue in 1887 - see \cite{Poincare}, {\bf 11}.} residue of the meromorphic 3-form
\begin{equation}
\label{omega3}
\omega_3:= \frac{d\xi_1\wedge d\xi_2\wedge d\xi_3}{f({\bf \xi})}, \, \,
f = \frac{1}{2}({\xi}^2 - \hbar^2 N^2)
\end{equation}
along the hypersurface $f=0$. The {\it Poincar\'e residue} of a meromorphic n-form
\begin{equation}
\label{omega/n}
\omega_n = \frac{g(z)}{f(z)} dz_1\wedge ...\wedge dz_n,
\end{equation}
where $f$ and $g$ are holomorphic functions, is defined as a holomorphic 
(n-1)-form on the hypersurface $f(z) = 0$ which possesses a local extension 
$\rho$ to ${\mathbb C}^n$ such that $\omega_n =\frac{df}{f}\wedge \rho$. If
$\frac{\partial f}{\partial z_j}|_{f=0} \neq 0$ in some neighbourhood $U$ of
a point of the hypersurface $f=0$, then
\begin{equation}
\label{Res}
{\rm Res} \, \omega_n = \left. g(z) (-1)^{j-1}\frac{d\xi_{1} \wedge ... \wedge d\xi_{j} \wedge ... \wedge d\xi_{n}}{\frac{\partial f}{\partial \xi_{j}}} \right| _{f=0} 
\end{equation}
in $\rm{U}$. 

{\it Exercise 3.2} Compute the residue of the 3-form (\ref{omega3}) in terms of the variables $\xi$ and in terms of the spherical angles $\theta, \varphi$,
\begin{eqnarray}
\xi_1 + i\xi_2 = 2 z_1 {\bar z}_2 = \hbar N sin\theta e^{-i\varphi}, &  \\  \nonumber
\xi_3 =z_1{\bar z}_1 - z_2 {\bar z}_2 = \hbar N cos\theta  \; \; \; \; \; \; \;  & (2z_1 z_2 = \hbar N sin\theta e^{i\alpha}) .  
\end{eqnarray}
Prove that the result coincides with the restriction of the form $\omega$ (\ref{hf}) to the sphere (\ref{zp}) (with $dN = 0$). ({\it Hint}: prove that $\omega$ can be written as 
\begin{equation}
\label{sph}
\omega = \frac{\hbar}{2} (dN\wedge (d\alpha -cos\theta d\varphi) + 
N sin\theta d\theta \wedge d\varphi) 
\end{equation}
in spherical coordinates.) 

Recalling that $N$ is by assumption a positive integer (so that $\hbar N$ belongs to the spectrum  
of the oscillator Hamiltonian $H_0$ (Eq. (\ref{osc})) for $n=2$ we conclude that the integral of 
the symplectic form of the 2-sphere is quantized:
\begin{equation}
\label{N}
\int \frac{{\rm Res} \, \omega_3}{4\pi\hbar} = \frac{N}{4\pi} \int_{{\mathbb S}^2} 
sin\theta d\theta \wedge d\varphi = N (= 1, 2, ...),
\end{equation}
thus reproducing the integrality of the second cohomology group.

The quantum counterpart of the gauge invariant variables $\xi_j$ are the components of 
the angular momentum; more precisely (cf. Sect. 1.3),
\begin{equation}
\label{Mj}
M_j = \frac{1}{2}a^*\sigma_j a \, \Rightarrow [M_3, M_\pm] = \pm \hbar M_\pm, \, [M_+, M_-] = 2\hbar M_3 
\end{equation}
where $M_\pm = M_1 \pm iM_2 = a^*\sigma_\pm a \, (M_+ = a_1^* a_2, \, M_- = a_2^* a_1)$.

\smallskip
 
\subsection{${\mathbb C}^2$ as a hyperk\"{a}hler manifold}

\begin{flushright}
{\it I then and there felt the galvanic circuit close; and the sparks which fell
 from it were the fundamental equations between $i, j$ and $k$...}  \\
                        W.R. Hamilton - {\it letter to P.G. Tait}, October 1858
\end{flushright} 
              
\smallskip

{\it Quaternions} provide the real 4-dimensional space ${\mathbb R}^4$ with a structure 
of a non-commutative normed star division algebra. We set
\begin{eqnarray}
\label{q}
q = q^0 + q^1 I + q^2 J + q^3 K, \, q^* = q^0 - q^1 I - 
q^2 J - q^3 K, \nonumber \\
\, I^2 = J^2 = K^2 = I J K = - 1 \, \Rightarrow qq^* = q^* q = |q|^2 = 
\sum_{\mu=0}^3 (q^\mu)^2.
\end{eqnarray}
{\footnotesize The imaginary quaternion units $I, J, K$ can be defined as operators (real matrices) $I_L, J_L, K_L$ in ${\mathbb R}^4$ which provide a real representation of 
the Lie algebra $su(2)$:
\newpage
\begin{equation}
\begin{array}{c}
\label{ijk}
I q = -q^1 + q^0 I - q^3 J + q^2 K, \,
J q = -q^2 + q^3 I + q^0 J - q^1 K, \, \\
K q = -q^3 - q^2 I + q^1 J + q^0 K \,  \\
\\
\Rightarrow
I_L =\left( 
\begin{array}{cc}
-\epsilon & {\bf 0} \\
{\bf 0} & -\epsilon
\end{array}
\right) = -{\bf 1}\otimes \epsilon , \;  
{\bf 1} =\left( 
\begin{array}{cc}
1 & 0 \\
0 & 1
\end{array}
\right)  ,\; {\bf 0} = \left( 
\begin{array}{cc}
0 & 0 \\
0 & 0
\end{array}
\right) , \\ \epsilon = \left( 
\begin{array}{cc}
0 & 1 \\
-1 & 0
\end{array}
\right) \left( = i \sigma_2 \right) ;   \\
J_L =\left( 
\begin{array}{cc}
{\bf 0} & -\sigma_3 \\
\sigma_3 & {\bf 0}
\end{array}
\right) = -\epsilon\otimes\sigma_3 , \, \sigma_3 =\left( 
\begin{array}{cc}
1 & 0 \\
0 & -1
\end{array}
\right) ; \; \\
K_L =\left( 
\begin{array}{cc}
{\bf 0} & -\sigma_1 \\
\sigma_1 & {\bf 0}
\end{array}
\right) =-\epsilon\otimes\sigma_1 , \sigma_1 =\left( 
\begin{array}{cc}
0 & 1 \\
1 & 0
\end{array}
\right) .
\end{array}
\end{equation}

The right multiplication by $I, J, K$ gives rise to another set of operators $I_R, J_R, K_R$  which commute with $I_L, J_L, K_L$; the resulting six operators 
generate the Lie algebra $so(4)\simeq su(2) \oplus su(2)$.} 

One can introduce a complex symplectic form in ${\mathbb C}^2 \sim{\mathbb R}^4$, setting
\begin{equation}
\label{Omega}
\Omega = \omega_J + i\omega_K, \, \, \omega_J = dz_1\wedge dz_2 - d{\bar z}_1\wedge d{\bar z}_2,
\, \omega_K = dz_1\wedge d{\bar z}_1 + dz_2\wedge d{\bar z}_2. 
\end{equation}
{\footnotesize Viewing $(dz_1, dz_2, d{\bar z}_1, d{\bar z}_2)$ as a basis in the (trivial) cotangent bundle 
on ${\mathbb R}^4$ we can write:}
\begin{equation}
\label{JK}
\begin{array}{c}
\omega_J =\frac{1}{2} (dz, d{\bar z})\wedge J \left( 
\begin{array}{c}
dz \\ d{\bar z}
\end{array}
\right) ,
\, J =\left( 
\begin{array}{cc}
\epsilon & {\bf 0} \\
{\bf 0} & -\epsilon
\end{array}
\right) = \sigma_3 \otimes \epsilon, \, dz = (dz_1, dz_2);\\
 \omega_K =\frac{1}{2} (dz, d{\bar z})\wedge K \left( 
\begin{array}{c}
dz \\ d{\bar z}
\end{array}
\right) , \, K =\left( 
\begin{array}{cc}
{\bf 0} & {\bf 1} \\
-{\bf 1} & {\bf 0}
\end{array}
\right) = \epsilon \otimes {\bf 1}.
\end{array}
\end{equation}
Here $\omega_J$ is a holomorphic form of type $(2, 0) + (0,2)$, $\omega_K$ is of
 type $(1,1)$ with respect to the complex structure defined by $K$. The form 
$\Omega$ (\ref{Omega}), on the other hand, is a holomorphic form of type 
$(2, 0)$ in the complexification ${\mathbb C}^4$ of ${\mathbb R}^4$ with respect
 to the complex structure $I$. This means that 
\begin{equation}
\label{hol}
\Omega(X, (1+i I)Y) = 0,  \, \, \forall X, Y \in T{\mathbb C}^4.
\end{equation}
{\it Exercise 3.3} Deduce (\ref{hol}) using the identity
\begin{equation}
\label{0div}
(J + i K)(1 + i I) = 0.
\end{equation}
Note that while the algebra of real quaternions $\mathbb H$
has no zero divisors the above example shows that its complexification admits
such divisors: none of the two factors in the lefthand side of (\ref{0div}) is 
zero while their product vanishes identically.

We observe that the form (\ref{Omega}) can be written in I-holomorphic coordinates as
a manifestly $(2, 0)$-form:
\begin{equation}
\label{Owz}
\Omega = dw\wedge dz \, \, {\rm for} \, w = z_1 - i{\bar z}_2, \, z = z_2 + i{\bar z}_1. 
\end{equation}

We are now prepared to give a general definition. A smooth manifold $M$ is called {\it hypercomplex} 
if its tangent bundle $TM$ is equipped with three (integrable) complex 
structures $I, J, K$ satisfying the quaternionic relation of (\ref{q}). If, in 
addition, $M$ is equipped with a {\it Riemannian metric} $g$ 
which is {\it K\"ahler} with respect to $I, J, K$, - i.e., if they are 
compatible with $g$ and satisfy 
\begin{equation}
\label{LC}
\bigtriangledown I = 0, \, \bigtriangledown J = 0 , \, \bigtriangledown K = 0,
\end{equation}
where $\bigtriangledown$ is the Levi-Civita connection, then the manifold $(M,I,J,K,g)$ is called 
{\it hyperk\"ahler}. This means that the {\it holonomy} of $\bigtriangledown$ lies inside the 
group $Sp(2n)(= Sp(n, {\mathbb H}))$ of quaternionic-Hermitian endomorphisms.

{\footnotesize  The converse is also true: a Riemannian manifold is hyperk\"ahler if and only if its 
holonomy is contained in $Sp(2n)$. This definition is standard in differential geometry. (In physics 
literature, one sometimes assumes that the holonomy of a hyperk\"ahler manifold is precisely $Sp(n)$, 
and not its proper subgroup. In mathematics, such hyperk\"ahler manifolds are called {\it simple 
hyperk\"ahler manifolds}. In algebraic geometry, the word ``hyperk\"ahler'' is essentially synonymous 
with ``holomorphically symplectic'', due to the famous Calabi-Yau theorem. The notion of a 
hyperk\"ahler manifold is of a relatively recent vintage: it has been introduced in 1978 (16 years after Bargmann's paper) by Eugenio Calabi.}

The above hyperk\"ahler space ${\mathbb C}^2$ is closely related to the regular adjoint orbit of $sl(2, {\mathbb C})$:
\begin{equation}
\label{det} 
-det \left( 
\begin{array}{cc}
a & b \\
c & -a
\end{array}
\right) = a^2 + bc = \lambda \neq 0.
\end{equation}
The hyperk\"ahler structure of (co)adjoint orbits of semisimple complex Lie groups and the associated Nahm's 
equation are being studied since over two decades - see \cite{Kr}, \cite{K96}, as well as the lectures
\cite{Bi} and references therein.

\bigskip

\section{Other approaches. From Weyl to Kontsevich}
\setcounter{equation}{0}
\renewcommand\theequation{\thesection.\arabic{equation}}

We are leaving out one of the most important topics of quantum theory: the path integral approach that would require another set
of lectures of a similar size. A 94-page preprint of such lecture notes is available \cite{Gr} with 93 references (up to 1992)
including the pioneer work of Dirac (1933) and Feynman\footnote{Richard Feynman (1918-1988) shared the Nobel Prize in Physics in 1965
with Julian Schwinger (1918-1994) and Sin-Itiro Tomonaga (1906-1979).} (1948). The book \cite{ZJ} is recommended as {\it a highly 
readable introduction to the subject}. For a recent development in this area - see \cite{W10}. 

We provide instead a brief historical introduction to deformation quantization starting in Sect. 4.1 with the forerunners of the
modern development. (Taking a more expansionist point of view and relating path integrals to star exponentials - see \cite{S98},
Sect. II.3.2.1 - one can pretend to incorporate the path integral approach into the vast domain of deformation quantization.)

\smallskip

\subsection{Quantum mechanics in phase space}

Prequantum mechanics lives in phase space - just like its classical antecedent. The polarization or the choice of a maximal set
of commuting observables, however, breaks, in a sense, the symmetry among phase space variables. Is that unavoidable? In 1927, in
the wake of the appearance of quantum mechanics and of the Heisenberg uncertainty relations, Hermann Weyl \cite{We} did propose 
a phase space formulation of quantization in which coordinates and momenta are treated on equal footing. Weyl maps any classical 
observable, i.e. any (smooth) function $f$ on phase space, to an operator $U[f]$ in a Hilbert 
space which provides a representation of the {\it Heisenberg-Weyl group} of the CCR. In the simplest case of a 2-dimensional
euclidean phase space with coordinates $(p, q)$ the Weyl transform reads:

\begin{equation}
\label{Weyl}
U [f] = \frac{1}{(2\pi)^2}\int ... \int f(q,p) e^{\frac{i}{\hbar}(a(Q-q) +b(P-p))} {\rm d}q\, {\rm d}p\, {\rm d}a\, {\rm d}b. 
\end{equation}
Here $P$ and $Q$ are the generators of the Heisenberg Lie algebra (satisfying the CCR) so that $g(a, b, c):= e^{i(\frac{aQ+bP}{\hbar} + c)}$ 
is an element of the corresponding Heisenberg-Weyl group (introduced by Weyl and associated by mathematicians with the name of Heisenberg)
satisfying the composition law
\begin{equation}
\label{compos}
g(a_1, b_1, c_1) g(a_2, b_2, c_2) = g(a_1 + a_2, b_1 + b_2, c_1 + c_2 + \frac{1}{2}(a_1 b_2 - a_2 b_1)).
\end{equation} 
Given any group representation $U [g]$, the operator $U [e^{i(\frac{aq+bp}{\hbar} + c)}]$ (\ref{Weyl}) will give the representation of the 
group element $g(a, b, c)$. 

The Weyl map may also be expressed in terms of the integral kernel matrix elements of the operator,
\begin{equation}
\label{kern}
    \left\langle x| U [f] |y \right\rangle = \int_{-\infty}^\infty {{\rm d}p\over h} ~e^{ip(x-y)/\hbar}~ f\left({x+y\over2},p\right) . 
\end{equation}
The inverse of the above Weyl map is the {\it Wigner map} \cite{W32}, which takes the operator Φ back to the original phase-space kernel function f,
\begin{equation}
\label{Wigner}
    f(q,p)= 2 \int_{-\infty}^\infty {\rm d}y~e^{-2ipy/\hbar}~ \left\langle q-y| U [f] |q+y \right\rangle. 
\end{equation}
The Wigner {\it quasi-probability distribution} in phase space corresponding to a pure state with wave function $\psi(x)$ is given by
\begin{equation}
\label{Wigpsi}
    F(x,p):= \frac{1}{\pi\hbar}\int_{-\infty}^\infty \psi^*(x+y)\psi(x-y)e^{2ipy/\hbar}\,dy\, 
\end{equation}
The qualification "quasi" is necessary since the distribution $F(x, p)$ may give rise to negative probabilities. We refer to \cite{M86}, \cite{Fe} (where more general non positive distributions are considered - see below) and to the entertaining historical survey \cite{CZ} for an explanation of how Heisenberg's uncertainty relation is reflected in the phase space formulation and prevents the appearance of physical paradoxes for an appropriate use of Wigner's distribution function. Someone, accustomed with the standard Hilbert space formalism of quantum mechanics, may still wonder why should one deal with such a strange formalism in which verifying a basic property like positivity of probabilities needs an intricate argument. Are there problems whose solution would motivate the use of the phase space picture?  Unexpectedly, a positive answer to this question has come from outside quantum mechanics. A little thought will tell us that, if we view q as a time coordinate and p as a frequency, then the Wigner function may serve to characterize a piece of music (or, more generally a sound signal) much better then having just a probability density of frequencies alone. Indeed, starting with the 1980's, applications of the Wigner distribution to signal processing has become an industry - see the monograph \cite{MH} and references therein. For an application to the decoherent history approach to quantum mechanics - see \cite{GH} and references to earlier work cited there. 

{\footnotesize The Wigner distribution (\ref{Wigpsi}) is real and has the property that if integrated in
either $p$ or $x$ it gives the standard quantum mechanical (positive) probability density with respect to the non-integrated variable; for instance,
\begin{equation}
\int F(x, p)dp = |\psi(x)|^2.
\end{equation}
Raymond Stora (private communication) has proposed another simple formula for the quasi-probability distribution $F = F_\rho$ corresponding to a (positive) density operator $\rho$ and a pair of (normalized) eigenstates $|\alpha>, |\beta>$ labelled by the eigenvalues of two (in general, non-commuting) hermitean operators which also satisfies these relations:
\begin{eqnarray}
\label{Stor}
F_\rho (\alpha, \beta) = \frac{1}{2}(<\alpha|\beta><\beta|\rho|\alpha> + <\alpha|\rho|\beta><\beta|\alpha>);
\sum_\beta F_\rho (\alpha, \beta) = <\alpha|\rho|\alpha>.
\end{eqnarray}
This formula applies to operators like spin projections on two orthogonal axes whose eigenvalues do not belong to an affine space, so that Wigner's expression (\ref{Wigner}) would not make sense. The appearance of negative probability is a common feature of all quasi-probability distributions consistent with Bell's theorem\footnote{For a ``probabilistic opposition'' to the ussual interpretation of Bell's theorem - see \cite{Kh}.} (as discussed in \cite{M86} and \cite{SR} among others). The first to consider negative probabilities (in the context of quantum theory) was none other than Dirac. In his Bakerian lecture \cite{D42} (p. 8) he stated ``Negative energies and probabilities should not be considered as nonesense. They are well defined concepts mathematically, like a negative sum of money...'' The Wigner transform has the extra property to be inverse to Weyl's which, in turn, is related to Weyl's (symmetric) ordering. There is, however, nothing sacred about such an ordering (or about any other ordering, for that matter). As mentioned earlier - see footnote 16 - Lax ordering naturally appears instead of Weyl's in the quantization of some integrable systems.}

Using the Weyl form (\ref{compos}) of the CCR and the Weyl correspondence von Neumann\footnote{The Hungarian born brilliant mathematician 
and polymath John von Neumann (1903-1957) made substantial contributions in a number of fields. In a short list of facts about his life he 
submitted to the National Academy of Sciences of the USA, he stated "The part of my work I consider most essential is that on quantum 
mechanics, which developed in G\"ottingen in 1926, and subsequently in Berlin in 1927-29. Also, my work on various forms of operator theory, 
Berlin 1930 and Princeton 1935-1939;" - see also \cite{V58}.} proved in 1931 \cite{vN31} (see also \cite{vN}, \cite{V58}) the essential uniqueness 
of the Schr\"odinger representation in Hilbert space. For completeness' sake, he worked out the image of operator multiplication discovering 
the convolution rule that defines the noncommutative composition of phase-space observables - an early version of what came to be called 
$\star$-product. In fact, once having the Weyl map $f\rightarrow U[f]$ and its inverse and knowing the operator product $U[f]U[g]$ we can 
define the star product $f\star g$ as the Wigner image of $U[f] U[g]$. The result is:
\begin{equation}
\label{star}
f\star g = \int \frac{dx_1 dp_1}{\pi\hbar} \int \frac{dx_2 dp_2}{\pi \hbar} f(x+x_1, p+p_1) g(x+x_2, p+p_2) exp(\frac{2i}{\hbar}(x_1p_2 - x_2 p_1)).
\end{equation}
   
{\footnotesize In fact, Weyl and Wigner introduced their maps for different purposes and neither of them noticed that they were inverse to each other 
or thought of defining a noncommutative star product in phase space. This was done independently by two young novices during World War II (see for more detail \cite{CZ}, \cite{ZFC}). 

The first was the Dutch physicist Hilbrand (Hip) Groenewold (1910-1996). After a visit to Cambridge to interact with John von Neumann (1934-5) on the links between classical and quantum mechanics, and a checkered career, working in Groningen, then Leiden, the Hague, De Bilt, and several addresses in the North of Holland during World War II, he earned his Ph.D. degree in 1946, under the Belgian physicist L\'eon Rosenfeld (1904-1974) at Utrecht University. Only in 1951 was he offered a position in theoretical physics at his Alma Mater in Groningen. It was his thesis paper \cite{G46} that laid the foundations of quantum mechanics in phase space. This treatise was the first to achieve full understanding of the Weyl correspondence as an invertible transform, rather than as an unsatisfactory quantization rule. Significantly, this work defined (and realized the importance of) the star-product, the cornerstone of this formulation of the theory, ironically often also associated with Moyal's name, even though it is not featured in Moyal's papers and was not fully understood by Moyal.
Moreover, Groenewold first understood and demonstrated that the Moyal bracket is isomorphic to the quantum commutator, and thus that the latter cannot be made to faithfully correspond to the Poisson bracket, as had been envisioned by Paul Dirac. This observation and his counterexamples contrasting Poisson brackets to commutators have been generalized and codified to what is now known as the Groenewold - Van Hove theorem. 

The second codiscoverer of the star product Jos\'e (Jo) Moyal (1910-1998) was born in Jerusalem, then in the the Ottoman Empire, and spent much of his youth in Palestine. After studying in France and Britain and working on turbulence and diffusion of gases in Paris, he escaped to London (with the help of the physicist/writer C.P. Snow (1905-1980)) at 
the time of the German invasion in 1940. While working on aircraft research at Hartfield, Moyal developed his ideas on the statistical nature of quantum mechanics and had an intense correspondence with Dirac\footnote{Feb 1944 - Jan 1946, reproduced in Ann Moyal {\it Maveric Mathematician} ANU E Press, 2006 (online).}, who refused to believe that there could be a "distribution function $F(p, q)$ which would give correctly the mean value of any $f(p, q)$" even after Moyal found out - and wrote to Dirac - that such a function was constructed by Wigner, Dirac's brother in law... Moyal eventually published his work in \cite{M49}, three years after Groenewold. Subsequent work on this topic, done during the next 15 years is reproduced in \cite{ZFC}. The subject only attracted wider attention another fifteen years later, after the work \cite{BFLS} triggered the interest of mathematicians to deformation quantization. Even then only references to Moyal surged dramatically while the work of Groenewold is still rarely mentioned (for instance, the paper \cite{G46} is not included among the 78 refernces of the 2008 survey \cite{B08} of deformation quantization).}

 \smallskip
 
\subsection{Deformation quantization of Poisson manifolds}

The natural starting point for the study of quantization is a {\it Poisson algebra} ${\mathcal A}$ - i.e., an associative 
algebra with a Poisson bracket that gives rise to a Lie algebra structure and acts as a derivation (obeying the Leibniz rule) 
on ${\mathcal A}$. In the case of a classical phase space this is the (commutattive) algebra of functions on a Poisson manifold.
The aim is to deform the commutative product to a $\hbar$ dependent noncommutative star ($\star$-)product in 
such a way that the star-commutator reproduces the Poisson bracket up to higher order terms in $\hbar$:
\begin{equation}
\label{starh}
f\star g - g\star f = i\hbar\{f, g\} + O(\hbar^2).
\end{equation}
{\it Deformation quantization} is computing and studying an {\it associative star product}, 
defined as a formal power series in $\hbar$: 
\begin{equation}
\label{starB}
f\star g = fg + \sum_{n=1}^\infty \hbar^n B_n(f, g),
\end{equation}
where $B_n$ are {\it bidifferential operators} (bilinear maps that are differential operators in each argument) and $B_1$ is
restricted by (\ref{starh}). Given the product (\ref{starB}) we can extend it to the algebra ${\mathcal A}[[\hbar]]$ of {\it 
formal power series} in the parameter $\hbar$ (with coefficients in ${\mathcal A}$) by bilinearity and {\it $\hbar$-adic continuity}:
\begin{equation}
(\sum_{n\geq 0}f_n\hbar^n)\star (\sum_{n\geq 0}g_n\hbar^n) = \sum_{k,l\geq 0, \, m\geq 1} B_m(f_k, g_l)\hbar^{k+l+m}.
\end{equation}
One considers \cite{W94} {\it gauge transformations} $G(\hbar): {\mathcal A}[[\hbar]] \rightarrow {\mathcal A}[[\hbar]]$ which preserve
the original Poisson algebra ${\mathcal A}$; in other words, $G(\hbar) = 1 + \sum_{n\geq 1} G_n \hbar^n$ where $G_n$ are (linear) 
differential operators. Two star products $\star$ and $\star^{'}$ are equivalent if they differ by a gauge transformation - i.e., if 
\begin{equation}
\label{equiv}
\sum_{j+k+l=n}B_l(G_j(f), G_k(g)) = \sum_{l+m=n} G_m(B^{'}_l(f, g)), \, \, n = 1, 2, ... . 
\end{equation}
The problem, stated in \cite{BFLS} (see also \cite{W94}), is to find (cohomological) conditions for existence of a star product and to classify all such products up to gauge equivalence.

First of all, we note, following \cite{K}, that the associativity of the star product implies the following relation for 
the first bidifferential operator $B_1$ of the series (\ref{starB}):
\begin{equation}
\label{cocyc}
fB_1(g, h) - B_1(fg, h) + B_1(f, gh) - B_1(f, g)h = 0.
\end{equation}
If we view $B_1$ as a linear map $B_1:  {\mathcal A}\otimes{\mathcal A} \rightarrow {\mathcal A}$ then Eq. (\ref{cocyc}) shows that it is 
a 2-cocycle of the cohomological Hochschild\footnote{Gerhard Hochschild (1915-2010), a student at Princeton of Claude Chevalley (1909-1984),
introduced the Hochschild cohomology in 1945.} complex of the algebra ${\mathcal A}$ (defined in Sect. 3.2.4 of \cite{K}). Furthermore, one can 
annihilate the symmetric part of $B_1$ by an appropriate gauge transformation (\cite{K} Sect. 1.2) thus ending up with 
$B_1(f, g) = \frac{i}{2}\{f, g\}$ - as a consequence of (\ref{starh}). More generally, in attempting to construct recursively $B_n$ one finds 
at each stage an equation of the form $\delta B_n = F_n$ where $F_n$ is a quadratic expression of the lower (previously determined) terms. A 
similar equation arises for each $G_n$ in the gauge equivalence problem. The operator $\delta$ goes from bilinear to trilinear (or from linear 
to bilinear) and is precisely the coboundary operator for the Hochschild cohomology with values in ${\mathcal A}$ of the algebra ${\mathcal A}$
(see \cite{W94}).

The simplest example of a star product is given by the Groenewold-Moyal product (\ref{star}), defined in terms of the Poisson bivector ${\mathcal P}$
(Sect. 2.1) with constant coefficients which exists in an affine phase space. It is given by (\ref{starB}) with
\begin{equation}
\label{GrMo}
B_n(f, g) = \frac{1}{n!}(\frac{i}{2}{\mathcal P}^{jk}\frac{\partial}{\partial y^j}\frac{\partial}{\partial z^k})^n f(y) g(z)|_{y=z=x}. 
\end{equation}

{\it Exercise 4.1} Verify (using (\ref{starh}), (\ref{starB}) and (\ref{GrMo})) the relation
\begin{equation}
\label{Wpq}
pq = \frac{1}{2}(p\star q + q\star p) = qp \, \, (q\star p - p\star q = i\hbar).
\end{equation}

{\footnotesize {\it Remark 4.1} In most mathematical texts, including \cite{K}, the i-factors in (\ref{starh}) and (\ref{GrMo}) are missing. (The Bourbaki 
seminar \cite{W94} is a happy exception. There the {\it parity condition} $B_n(f, g) = (-1)^n {\bar B_n(g, f)}$ which uses complex conjugation is also 
mentioned.) To make formulas conform with physics texts one has to substitute the formal expansion parameter by $i\hbar$. If a 
similar discrepancy in the writings of some of the founding fathers of geometric quantization could be viewed as a negligence, in the case of 
deformation quantization it seems to be a deliberate choice. Kontsevich is making the following somewhat criptic {\it Remark 1.5} in \cite{K}:
{\it In general, one should consider bidifferential operators with complex coefficients ... . In this paper we deal with purely formal algebraic properties
... and work mainly over the field ${\mathbb R}$ of real numbers. ... it is not clear whether the natural physical counterpart for the 'deformation quantization' 
for general Poisson brackets is the usual quantum mechanics. It is definitely the case for nondegenerate brackets, i.e. for symplectic manifolds, but our results 
show that in general a topological open string theory is more relevant.}}    

If the Poisson bivector ${\mathcal P}$ has a constant rank, then according to a classical theorem by Lie (cited in \cite{W94}) the Poisson manifold is locally 
isomorphic to a vector space with constant Poisson structure. Such {\it regular Poisson manifolds} are, hence, {\it locally} deformation quantizable. The local 
quantization can be patched together relatively easily if there exists a torsionless linear connection such that  ${\mathcal P}$ is covariantly constant \cite{BFLS}. 
The more difficult problem to prove existence of deformation quantization for arbitrary symplectic manifolds which do not admit flat torsionless Poisson connections has 
been solved by de Wilde and Lecomte and by Fedosov in the 1980's (see for reviews \cite{W94} and \cite{B08}). Weinstein ends his Bourbaki seminar talk \cite{W94} by 
asking the fundamental question "{\it Is every Poisson manifold deformation quantizable?}". Three years later, Kontsevich \cite{K} not only gave an affirmative 
answer to this question but provided a canonical construction of an equivalence class of star products for any Poisson manifold. This result was cited among his 
"contributions to four problems of geometry" for which he was awarded the Fields Medal in Berlin in 1998. The quantum field theoretic roots of this work were 
displayed in a series of papers of Cattaneo and Felder (see \cite{CF} and earlier work cited there).

As stressed in \cite{GW}, the convergence problem for the formal power series involved in the star-product 
is still only studied on a case by case basis.

The vitality of the subject is witnessed by a continuing flow of interesting papers - see e.g. \cite{C07}, \cite{CFR}, \cite{LW} among many others.
\bigskip

\section{Second quantization}
\setcounter{equation}{0}
\renewcommand\theequation{\thesection.\arabic{equation}}

\hfill\begin{minipage}{.7\linewidth}
{\it ... nothing gives greater pleasure to the conoisseur, ... even if he is a historian contemplating it retrospectively,
accompanied nevertheless by a touch of melancholy. The pleasure comes from the illusion and from the far from clear meaning;
once the illusion is dissipated, and knowledge obtained, one becomes indifferent... a theory whose majestic beauty no longer 
excites us.} A 1940 letter of Andr\'e Weil (from a French prison, to his sister Simone) on analogy in mathematics, 
{\it Notices of the AMS} {\bf 52}:3 (2005) 334-341 (p.339).
\end{minipage}

\smallskip

The story of inventing second quantization is the story of understanding 
``quantized matter waves'', and ultimately, of creating quantum field theory. 
Following its early stages (with a guide like \cite{Dar}) one may appreciate 
the philosophical inclinations of the founding fathers which appear to be no 
longer in the spirit of our days. It may also shed light on some of our current
 worries (\cite{S10}). But, first of all, we realize how difficult it has been 
to come to terms with some ideas which now appear as a commonplace. One such
idea, put forward by Jordan in 1927 (following a vague suggestion by Pauli - see
 \cite{Dar}, p. 230, footnotes 75. and 76.) and worked out in a final form by 
Jordan and Wigner\footnote{Jeno (later Eugene) Wigner (Budapest, 1902 - 
Princeton, 1995) was awarded relatively late (in 1963) the Nobel Prize in 
Physics ``for ... the discovery and application of fundamental symmetry 
principles''.} by the end of the year (l.c. pp. 231-232 and \cite{JW28}), was 
the introduction of the canonical anticommutation relations
\begin{equation}
\label{CAR}
[a_i, a_j]_+:= a_i a_j + a_j a_i=0 = [a_i^*, a_j^*]_+\,, \ \ [a_i, a_j^*]_+ = 
\delta_{ij}
\end{equation}
as a basis of Fermi statistics (and indeed of the quantization of the electron-positron field). The difficulty in accepting the canonical anticommutation relation stems from the fact that they seem to violate the correspondence principle: for $\hbar \rightarrow 0$ they become strictly anticommuting (Grassmann) variables, never encountered 
before in a classical system.  It was natural for Jordan to coin the term {\it second quantization} since he was quantizing the (already quantum) Schr\"odinger wave 
function (see Appendix). Dirac, on the other hand, was concerned with quantizing a classical system: the electromagnetic radiation field, \cite{D27}. This should help us understand why even he, the codiscoverer of the Fermi-Dirac statistics, was not ready to accept such a notion. In the Solvay congress of 1927 he ``argues  that [the Jordan-Wigner's quantization] is very artificial from a general point of view.'' (see \cite{Dar}, p. 239). Nearly half a century later Dirac remembers: ``Bose statistics ... was connected ... to an assembly of oscillators. There was no such picture available with the Fermi statistics, and I felt that was a serious drawback.'' 
(see \cite{D}, p. 140). Had Dirac applied the canonical anticommutation 
relations to his wonderful relativistic wave equation, he would not have needed
 the ``filled up infinite sea\footnote{A precise mathematical formulation of the
 {\it Dirac sea}, equivalent to the now standard quantum theory of a Fermi 
field, has been only given recently, \cite{D11}.} of negative-energy states''. 
Jordan in fact anticipated the {\it spin statistics theorem} (which states that
 integer spin fields locally commute while half-integer spin fields locally 
anticommute) formulated and proven some 12 years later by Markus Fierz 
(1912-2006) and Pauli (for a pedagogical discussion of this theorem, its history
 and understanding - see the 20-page-long paper \cite{DS}, available 
electronically, that contains over fifty original references).

The mathematical formulation of second quantization is clean and elegant (and, 
in the spirit of the above cited letter of Weil, hides much of the excitement).
 Second quantization, in the narrow sense of quantizing the Schr\"odinger wave 
function, can be viewed as an attempt to get a quantum description of a 
many-particle system from the quantum description of a single particle. Starting
 from a single particle Hilbert space $\mathcal{H}$ one forms the symmetric (or
 antisymmetric) tensor algebra $S(\mathcal{H})$ (or $A(\mathcal{H})$) and 
completes it to form a bosonic (or fermionic) {\it Fock space} \footnote{The 
Saint Petersburg's physicist Vladimir Fock (1898-1974) is also known for his 
development of the Hartree-Fock method and its relativistic counterpart, the 
Dirac-Fock equations, which led to the work of Dirac-Fock-Podolsky on quantum 
field theory, a precursor of Tomonaga's Nobel prize winning formulation 
involving infinitely many times. His ground-breaking work \cite{F32}, duly cited
 in the students' paper \cite{CF09}, is oddly absent from the list of references
 of the major treatise \cite{Sch}.} $\mathcal{F} = \mathcal{F}(\mathcal{H})$. 
More generally, one has a functor, called {\it second quantization} from the
Hilbert category to itself, which sends each Hilbert space to its Fock space, 
and each unitary operator $U$ to an obvious
unitary map (built out of tensor products of $U$'s).

Here is a toy example of a bosonic Fock space presented by Bernard Julia to the 1989 Les Houches Winter School on
Number Theory and Physics, which served a starting point of an interesting
mathematical development \cite{BC} (which is still continuing, \cite{CC}).
One introduces (Bose) creation and annihilation operators $a_p^{(*)}$, corresponding to the prime numbers, $p=2, 3, 5, \\
 7, 11, ...$. The 1-particle state space is spanned by (unit) vectors $|p>$ corresponding to the primes while the Fock 
space $\mathcal{F}$ is spanned by vectors $|n>$ corresponding to all positive integers:
\begin{equation}
\label{JBC}
|vac>\equiv |1>\,, \ \ |n> = \frac{\prod(a_i^*)^{n_i}}{\prod(n_i!)^{\frac{1}{2}}}
|1> \ \ {\rm for} \ \ n = \prod (p_i)^{n_i}\,.
\end{equation}
Thus the vacuum corresponds to the number 1; the states $|4=2^2>, |6=2\times 3>, |9>, |14>, ...$ are 2-particle states etc.
The {\it number operator} $N$ such that $(N-n)|n> = 0$ acts multiplicatively on product states:
\begin{equation}
\label{N}
N:= \prod_p p^{a^*_p a_p} \ \ \Rightarrow N|p_1...p_k> = p_1...p_k |p_1...p_k>\,.
\end{equation}

If one introduces furthermore a logarithmic Hamiltonian that is additive on 
product states then the partition function of the system, corresponding to 
inverse temperature $\beta$, will be the Riemann zeta-function:
\begin{equation}
\label{HZ}
H = ln N \ \ \Rightarrow \ \ Z(\beta):= tr_{\mathcal F}(e^{-\beta H}) = \sum_1^\infty \frac{1}{n^\beta} =\zeta(\beta)=
\prod_p(1-\frac{1}{p^\beta})^{-1}\,.
\end{equation}

{\it Remark 5.1} Rather than using symmetrized or antisymmetrized tensor products of 1-particle spaces we could use higher
dimensional irreducible representations of the permutation group corresponding to more general {\it permutation group or
parastatistics} which appear in the classification of supersellection sectors in the algebraic Doplicher-Haag-Roberts
approach to local quantum physics (for a review - see \cite{H}). They can be reduced to the familiar Bose and Fermi
statistics (by the so called {\it Green ansatz}) at the expense of introducing some extra degrees of freedom and a
{\it gauge symmetry}.

The Fock space construction works nicely for free quantum fields as well as in nonrelativistic quantum mechanics, whenever the Hamiltonian commutes with the particle number. The tensor product construction is not appropriate even for treating the nonrelativistic bound state problem. Consider, indeed, the tensor product of the state spaces of two Galilean invariant particles. According to a classical paper by Bargmann \cite{B54} the quantum mechanical ray representation of the Galilean group involves its central extension by the mass operator. Thus the mass of the tensor product of two 1-particle representations, equals to the sum of the masses of the two particles, should be conserved. On the other hand, we know that the mass of a bound state differs from the sum of the constituent masses by the (negative) binding energy \footnote{This remark continues a discussion provoked by a Scholarpedia article on the subject.}(divided by $c^2$). A similar contradiction is reached by considering the energy conservation implied by the Galilean invariance of the tensor product. This example suggests that in the presence of interactions one should consider a nontrivial {\it coproduct}, such that a symmetry generator like the total energy is not necessarily additive. Although the idea of a Hopf algebra deformation of second quantization has been explored by a number of authors (see e.g. \cite{CCT}), I am not aware of a work addressing the physical bound-state problem in this manner.

{\bf Acknowledgments}. I greatfully acknowledge the hospitality of IHES (in January and in December 2011), and of the Theory Division of CERN (in February and March, 2012) during the course of this work. It is a pleasure to thank Peter Dalakov, Petko Nikolov and Raymond Stora for useful suggestions at different stages of this work and Hristina Hristova for her help in preparing the notes. The author's work has been supported in part by grant DO 02-257 of the Bulgarian National Science Foundation.
\newpage

\section*{Appendix. Pascual Jordan (1902-1980)}

\hfill\begin{minipage}{.7\linewidth}
{\it Among the creators of quantum mechanics Pascual Jordan is certainly the least known, although
he contributed more than anybody else to the birth of quantum field theory.}
Olivier Darrigol \cite{Dar}
\end{minipage}

\bigskip

Born in Hannover in a mixed German-Spanish family, well read in the natural sciences,
Pascual Jordan dreamed at the age of 14 to write ``a big book about all fields of science''.
He taught himself calculus while in the Gymnasium and ended up with a careful study of Mach's
{\it Mechanik} and {\it Prinzipien der W\"armelehre}.\footnote{The Austrian physicist and philosopher Ernst Mach (1838-1916)
had strong antimetaphysical views that influenced his godson Pauli (as well as the young Einstein). Throughout his life 
Jordan considered himself a disciple of Mach and referred to his positivistic theory of knowledge \cite{Dar}. (Other
sources on P. Jordan: \cite{PJ07}, \cite{Sch99}, \cite{Me}, \cite{S06}.)} Not satisfied with the teaching of physics at the {\it
Technische Hohschule} in Hannover he moved to G\"ottingen in 1923. The (experimental) physics
lectures there being too early in the morning, he recalled (in an interview with
T.S. Kuhn in 1963) to have become a physicist ``who never attended a course of lectures on physics''.
By contrast he became an active student of Richard Courant (1888-1972) and assisted him in writing
parts of the famous Courant-Hilbert's book on Methods of Mathematical Physics. Jordan only decided
that he will pursue physics (rather than mathematics) after he met Max Born (1882-1970), the newly
appointed director of the Institute of Theoretical Physics in G\"ottingen. ``He was ... the
person who, next to my parents, exerted the deepest, longest lasting influence on my life.'',
wrote Jordan in a brief eulogy after Born's death (\cite {Sch}, p. 7). In the beginning he
was just helping his teacher by inserting formulas in the manuscript of Born's Encyclopaedia article on the dynamics of crystal lattices (see \cite{MR}, footnote 60), but soon he started working on his own on the then hot topic of light quanta (starting with his thesis of 1924).
In early 1925 he was able to predict the existence of two new spectral lines in neon (to be soon observed by Hertz\footnote{Gustav Ludwig 
Hertz (1887-1975), Nobel Prize in Physics, 1925 (with James Frank), is a nephew of Heinrich Hertz (1857-94), the discoverer of the
electromagnetic waves.} - these were times fecund in new discoveries!). 

Jordan's activity during the years 1925-28 was truly remarkable: while Born was on vacation he wrote the first 
draft of their article (submitted two months after Heisenberg's). Then came the famous ``three-man-paper'' with 
Born and Heisenberg, submitted in November, in which Jordan was the sole responsible for the part devoted to the 
radiation theory. As if that was not enough, by the end of the year he submitted a paper on the ``Pauli statistics''; 
Max Born, an editor of Zeitschrift f\"ur Physik, took it with him on his way to the United States for a lecture tour 
and ... forgot all about it until his return to G\"ottingen six months later. In the meantime, its result was discovered 
independently by Fermi and by Dirac\footnote{In the words of Stanely Deser, cited in \cite{S06}, we might have spoken 
about {\it Jordanons} instead of fermions... Jordan himself used the term "Pauli statistics". A half a century older Jordan 
\cite{J} recalls that ``in early discussions [Pauli] rejected the obvious idea of extending the scope of his law. Later,
as part of the Fermi-Dirac statistics, it attained the status of a ... fundamental law in physics.'' (No allusion to 
his priority!)}. In the bibliography given in \cite{PJ07} 
(pp. 175-206) one finds 8 titles (including a book) with the participation of Jordan, published in 1926, 15 in 1927, 6 in 1928. 
Two of them are concerned with the {\it transformation theory} (one of 1926 and another of 1927, written in a friendly 
competition with Dirac - whom he thanks in the printed version for mailing him his manuscript\footnote{As noted by Schroer \cite{S06},
there is a third nearly forgotten contributor to this subject, Fritz London (1900-1954), better known for his study of the hydrogen 
molecule and the superconductivity; London was the first to introduce, in 1926, the concept of a Hilbert space in quantum mechanics.}). 
This work laid in effect the mathematical and physical foundations of quantum mechanics. Five other papers, the first two by Jordan alone 
\cite{J27}, the remaining three - with Oskar Klein (1894-1977) \cite{JK27}, with Wigner (see footnote 33) \cite{JW28} (on the canonical 
anticommutation relations for fermions), and finally, with Pauli - also of 1928, are concerned with the concept of second quantization, 
or in other words, with the quantizaion of wave fields, thus laying the ground of quantum field theory. It is difficult nowadays to fully 
appreciate the novelty and the significance of this work. Why, for instance, should one quantize the wave function of the already quantum
Schr\"odinger equation? Here is an unexpected for us reason. A problem that still worried physicists in the late 1920's was the physical 
interpretation of the wave function. Schr\"odinger was trying, in 1926, to give a realistic physical meaning to his waves, to think of 
their modulus square, $|\psi|^2$, as a kind of density of electronic matter (\cite{Dar}, p. 237). One of the obstacles to such an 
interpretation (raised by the expert critic Pauli) was the necessity to introduce a multi-dimensional configuration space to deal with 
several-body problems. Regarding $\psi$ as a field operator, Jordan restored in a way the 3-dimensional picture for treating 
an arbitrary (even a changing) number of particles. Furthermore, Jordan and Klein \cite{JK27} 
were happy to discover that normal ordering in the operator formalism allowed to eliminate in a natural way the
infinite self-energy terms (\cite{Dar}, pp. 234-235). (The even more revolutionary fermionic second quantization
and its uneasy reception was discussed in Sect.  5.)

So why did not Jordan share the fame of his G\"ottingen colleagues? Not only he did not get a Nobel
Prize (in spite of the fact that the authors of the ``Dreim\"annerarbeit'' were proposed twice to the
Nobel committee by Einstein during the late 1920's \cite{S06}), he was the only major contributor to
the development of quantum theory who did not attend the glorious 1927 Solvay conference (17 of whose
29 participants were or became Nobel laureats - see \cite{Sch}, p. 6); during the 35 years he lived
after the War he was all but forgotten. The reasons for such a neglect are complex: they concern
Jordan's personality and politics (and reflect the fact that our society praises scientists not just
for their scientific achievements).

To begin with, it has not been easy for the twenty-year-old newcomer to G\"ottingen to withstand
the brash and confident ways of his brilliant one or two years older colleagues, Heisenberg and Pauli.
According to Schweber, \cite{Sch} p. 7, ``Jordan was rather short and his presentation of self
reflected his physical stature.'' Besides, he badly stuttered, this made it difficult for him to
communicate with others and reinforced the impression of insecurity which he left. The fact that he
had affinity for mathematical problems and techniques (including the study of {\it Jordan algebras}\footnote{It is a non-associative 
algebra characterized by the relation $A^2\circ(A\circ B) = A\circ(A^2\circ B) \, (A^2 = A\circ A)$ satisfied by the symmetric 
product $A\circ B := \frac{1}{2}(A B + B A)$.} to which a joint work with von Neumann and Wigner \cite{JNW} (of 1934) is devoted) 
did not enhance his popularity among physicists\footnote{His post-war student Engelbert Sch\"ucking, \cite{Sch99}, 
recounts: ``Jordan was looked down upon  by Pauli and Heisenberg as more of a mathematician than 
a physicist'', and "Herr Jordan was always a formalist", Pauli once told me. Jordan, by contrast,
has only praise for Pauli - see his insightful essay \cite{J}.} (or with the Nobel committee, for 
that matter: even the great Poincar\'e (was nominated for but) did not receive the Nobel Prize). 
As observed by Freeman Dyson, one has to stick long enough to the field of his greatest success, 
if his aim is to get a Nobel Prize. By contrast, faithful to the dream of his 14-year-old self
to embrace the whole of science, Jordan moved on in the 1930's to problems in biology,
psychology, geology, and cosmology. He was one of the very first scientists who subscribed
before World War II to the big-bang hypothesis\footnote{His cosmological ideas followed the theory of
the Belgian priest and astronomer Monsignor Georges Lema\^itre (1894-1966) whose discovery of the redshift-distance
relationship was later ascribed to Edwin Hubble (1889-1953) - see \cite{WN}. They were also inspired by Dirac's 1937
large number hypothesis). Jordan's contributions were rediscovered and became popular (without crediting their
originator) decades later - see H. Kragh in \cite{PJ07}.}. If, thus, in the late twenties and early
thirties the lack of full recognition may be traced to Jordan's insufficient self-assertiveness and
his uncommonly wide interests, the way he was ignored after the War has to do with his politics.

The resentment against the humiliating Versailles treaty and the economic hardship aggravated by
exorbitant reparations were a fertile soil for the springing of nationalist feelings and for the
rise of political extremism. To cite once more Schucking \cite{Sch99}: "Jordan had been a conservative nationalist who published his 
elitists views in the right wing journal {\it Deutsche Volkstum} (German Heritage) under the pseudonym 'Domeier'. My 
G\"ottingen teacher Hans Kopfermann ... wrote to Niels Bohr in May 1933: 'There is a tendency among the non-Jewish younger scientists 
to join the movement and to act as much as possible as a moderating element, instead of standing disapprovingly on the sidelines'."
Indeed, Jordan was among the 8.5 million Germans to join the National-Socialist (NS) Party after Hitler came to power; he even took part 
in its semimilitary wing SA (the Storm Troopers or ``brown shirts'' who became largely irrelevant after the "Blood purge" of 1934 against their leaders). 
Much like the last liberal British Prime Minister Lloyd George (see \cite{CMM}), Jordan thought that the spread of communism from Soviet 
Russia was the greatest danger and a national-socialist Germany was the only alternative. Bert Schroer shares
in \cite{PJ07} the above cited argument that Jordan had the naive hope to convince some influential people in the NS
establishment that modern physics, as represented by Einstein and especially by
the new Copenhagen version of quantum physics, was the best antidote against
``the materialism of the Bolsheviks''. This view is corroborated by Jordan's
book \cite{J36} which is inspired (and refers approvingly to) Bernhard Bavink\footnote{German physics
teacher, philosopher of science and prolific author (1879-1947).}. Bavink argues that modern physics
was thoroughly anti-materialistic and in far better agreement with Christian belief than classical
physics (see H. Kragh in \cite{PJ07} and references therein). Not surprisingly, such views were not
welcome by the Nazi authorities, obsessed, as they were, by antisemitism. They accepted Jordan's support
but never trusted him as he continued his association with (and was ready to publicly praise) Jewish
colleagues. He spent some 16 years, 1928-1944 in a relative isolation, at the small University of Rostock
and was never assigned an important war related task (as was, for instance, Heisenberg who did not join the
party). Jordan did only inflict harm on his own reputation: for two years after the war he did not have any
work. Born refused to witness on his behalf, citing (in a letter responding to his request) the names of his relatives who perished 
during the Nazi rule (see \cite{B05}). Jordan only passed eventually the process of denazification with the help of Heisenberg and 
Pauli (and had to wait until 1953\footnote{The nationalist philosopher Theodor Haering (1884-1964), whose 
obscurantist views on modern physics Jordan criticized in his book \cite{J41} during the Nazi time, was rehabilitated 
two years earlier, in 1951. (I thank K.-H. Rehren for this information.)}  to be allowed to advise PhD candidates). Once reinstated 
as a professor at the University of Hamburg, he created a strong school of general relativity\footnote{His students included
J\"urgen Ehlers (1929-2008), who became in 1995 the founding director of the newly created Max Planck Institute for Gravitational 
Physics (Albert Einstein Institute) in Golm (Potsdam), and Sch\"ucking who ended his career in the New York University. In the
hands of the Hamburg group Dirac's idea of a variable gravitational constant was transformed into the still popular scalar-tensor
theory of gravity, usually attributed to Brans-Dicke (who wrote their paper in 1961, two years after Jordan).}(see \cite{E09}). But 
Jordan did not follow Pauli's advice to stay away from politics. Opposing the manifesto of the ``G\"ottingen eighteen" 
(of April 1957, signed by Born and Heisenberg) - against the nuclear rearmament of Germany - he wrote a 
counter article in support of Adenauer's policy claiming that the action of the eighteen endangered world
peace and undermined the stability in Europe. Max Born was irritated by Jordan's article but did
not react in public. (His wife did not hide her anger: she collected and published Jordan's old political
articles under the title ``Pascual Jordan, propagandist on the pay of CDU''.)

Eugene Wigner (Nobel Prize in Physics of 1963) nominated in 1979 (from Princeton) his former
colleague (and coauthor of \cite{JW28}) for the Nobel Prize, but to no avail: that year the Nobel
Prize in Physics was shared among Sheldon Glashow, Abdus Salam and Steven Weinberg - ``three
practitioners of the art that Jordan had invented'', in the words of Schucking \cite{Sch99}.

Pascual Jordan died on July 31, 1980 in Hamburg, three months before reaching 78, 
still working on his scalar-tensor theory of gravity.

\bigskip


\end{document}